\documentclass[aps,prx,superscriptaddress,amsmath,amssymb,
floatfix,twocolumn]{revtex4-1}
\usepackage{graphicx}
\usepackage{subfigure}
\usepackage{color}
\usepackage{epstopdf}
\usepackage{booktabs}
\usepackage{multirow}
\newcommand{\ry}[1]{\textcolor{black}{#1}}

\begin{document}
	\title{Spin excitations of the Shastry-Sutherland model -- altermagnetism and
		\ry{deconfined quantum criticality}}
	\author{Hongyu Chen}
	\thanks{These authors contributed equally to the work.}
	\affiliation{School of Physics and Beijing Key Laboratory of Opto-electronic
		Functional Materials and Micro-nano Devices, Renmin University of China,
		Beijing 100872, China }
	
	\author{Guijing Duan}
	\thanks{These authors contributed equally to the work.}
	\affiliation{School of Physics and Beijing Key Laboratory of Opto-electronic
		Functional Materials and Micro-nano Devices, Renmin University of China,
		Beijing 100872, China }
	
	\author{Changle Liu}
	\altaffiliation{liuchangle89@gmail.com}
	\affiliation{School of Engineering, Dali University, Dali, Yunnan 671003,
		China}
	
	\author{Yi Cui}
	\affiliation{School of Physics and Beijing Key Laboratory of Opto-electronic
		Functional Materials and Micro-nano Devices, Renmin University of China,
		Beijing 100872, China }
	\affiliation{Key Laboratory of Quantum State Construction and Manipulation
		(Ministry of Education), Renmin University of China, Beijing, 100872, China}
	
	\author{Weiqiang Yu}
	\affiliation{School of Physics and Beijing Key Laboratory of Opto-electronic
		Functional Materials and Micro-nano Devices, Renmin University of
		China, Beijing 100872, China }
	\affiliation{Key Laboratory of Quantum State Construction and Manipulation
		(Ministry of Education), Renmin University of China, Beijing, 100872, China}
	
	\author{Z. Y. Xie}
	\altaffiliation{qingtaoxie@ruc.edu.cn}
	\affiliation{School of Physics and Beijing Key Laboratory of Opto-electronic
		Functional Materials and Micro-nano Devices, Renmin University of
		China, Beijing 100872, China }
	\affiliation{Key Laboratory of Quantum State Construction and Manipulation
		(Ministry of Education), Renmin University of China, Beijing, 100872, China}
	
	\author{Rong Yu}
	\altaffiliation{rong.yu@ruc.edu.cn}
	\affiliation{School of Physics and Beijing Key Laboratory of Opto-electronic
		Functional Materials and Micro-nano Devices, Renmin University of
		China, Beijing 100872, China }
	\affiliation{Key Laboratory of Quantum State Construction and Manipulation
		(Ministry of Education), Renmin University of China, Beijing, 100872, China}
	
	\begin{abstract}
		\ry{Frustrated quantum magnets can host a variety of exotic spin 
		excitations, including fractionalized spin excitations coupled 
		to emergent gauge fields at deconfined 
		quantum critical points (DQCPs) and chiral magnons in altermagnets. 
		Here, we investigate the spin excitation spectra of the highly 
		frustrated $S=1/2$ antiferromagnetic (AFM) Shastry-Sutherland model, 
		focusing on the evolution of low-energy collective modes from the 
		N\'{e}el AFM 
		phase to the plaquette valence bond solid (PVBS). We 
		demonstrate that the AFM state exhibits altermagnetic 
		behavior, 
		characterized by a non-relativistic splitting between two chiral magnon 
		bands. Furthermore, we identify two additional low-energy excitations: 
		a Higgs mode in the longitudinal excitation channel 
		and an $S=0$ excitation with vanishing spectral weight. As the system 
		approaches the AFM-to-PVBS transition, 
		both these modes soften along with the lowest-energy triplet and 
		singlet modes in the PVBS state. The closing gap of the Higgs mode, 
		combined with the nearly degenerate velocities of $S=1$ and $S=0$ 
		excitations, provides spectral evidence that the AFM-to-PVBS transition 
		is proximate to a DQCP with emergent 
		$O(4)$ symmetry. Our results help clarify the spectral signature of 
		a broad class of symmetry enhanced quantum phase transitions 
		including   
		deconfined quantum criticality.
		}
	\end{abstract}
	\maketitle
	
	
	\section{Introduction}
	Symmetry 
	provides an
	organizational principle to classify complex states of matter, and based on
	this, many exotic phenomena are understood~\cite{TI_Review,
		Senthil_AnnRevCondMattPhys_2015,
		Witten_NP_2018}. One
	prominent
	example is
	the recently proposed altermagnet, a collinear antiferromagnetic (AFM)
	state in which
	the two sublattices with oppositely polarized spins are connected by the
	rotational symmetry instead of inversion or
	translation~\cite{Smejkal_PRX_2022A,Smejkal_PRX_2022B}. The
	distinct symmetry,
	which is organized by the spin space group theory~\cite{Yang_arXiv:2105.12738},
	causes a
	non-relativistic spin-dependent splitting of energy bands and may give rise to
	some exotic effects, such as anomalous Hall
	effect~\cite{Smejkal_PRX_2022A,Smejkal_NatRevMater7_2022}. Up to
	now, \ry{significant progress has been made 
	in the study of} itinerant systems~\cite{Gonzalez_Betancourt_PRL_2023,
		Hariki_PRL_2024, Ferrari_arXiv:2408.00841, Wu_PRB_2007, Liu_NC_2021, 
		Satoru_JPSJ_2019}. For insulating 
		altermagnets, though
	a symmetry dictated splitting in magnon
	bands has been predicted~\cite{Smejkal_PRL131_256703_2023}, existing
	studies are mostly limited to several model systems~\cite{Ma_PRB110_064426_2024,
		Weissenhofer_PRB110_094427_2024, Yao_arXiv_2024, Zhaojize_arxiv}.
	
	Symmetry can also dictate phase transitions. In the
	Landau-Ginzburg-Wilson paradigm, a system transitions from a
	disordered state to an ordered one by spontaneously
	breaking a symmetry. With quantum fluctuations the transition can be continuous
	and occur at zero temperature, giving rise to intriguing quantum critical
	phenomena~\cite{sachdev_2011,Coldea_2010,Senthil2004,Cui2023DQCP,Cui2019SCVO,
		Lee2019,E8_2021,Xu2022,agrawal2020quantum,fey2019quantum,zhang2011exploring}. For a transition between two ordered phases,
	the Landau paradigm predicts a first-order transition. However, this scenario
	has
	recently been challenged. It was proposed that the transition between a valence
	bond solid (VBS) and an antiferromagnetic (AFM) phase can be continuous,
	\emph{e.g.}, via a deconfined quantum critical point (DQCP)~\cite{Senthil2004}.
	At the DQCP, the establishment of one order is accompanied by the destruction
	of the
	other order via proliferation of topological excitations, giving rise to
	emergent deconfined fractionalized excitations as well as
	enhanced continuous symmetry among the seemingly unrelated two types of order
	parameters.
	This scenario has motivated extensive studies on
	DQCP~\cite{Sandvik2007,
		Meng_EasyplaneDQCP, Shao_Science, Senthil_Vojta_Sachdev:2004, Liu_Vojta:2022,
		Guo_PRL_2024}.
	However, whether a DQCP can be realized in
	quantum spin models \ry{and what would be its spectral signatures  
	are 
	still 
	challenging questions.}
	
\begin{figure}[t]
		\centering
		\includegraphics[width=0.95\linewidth]{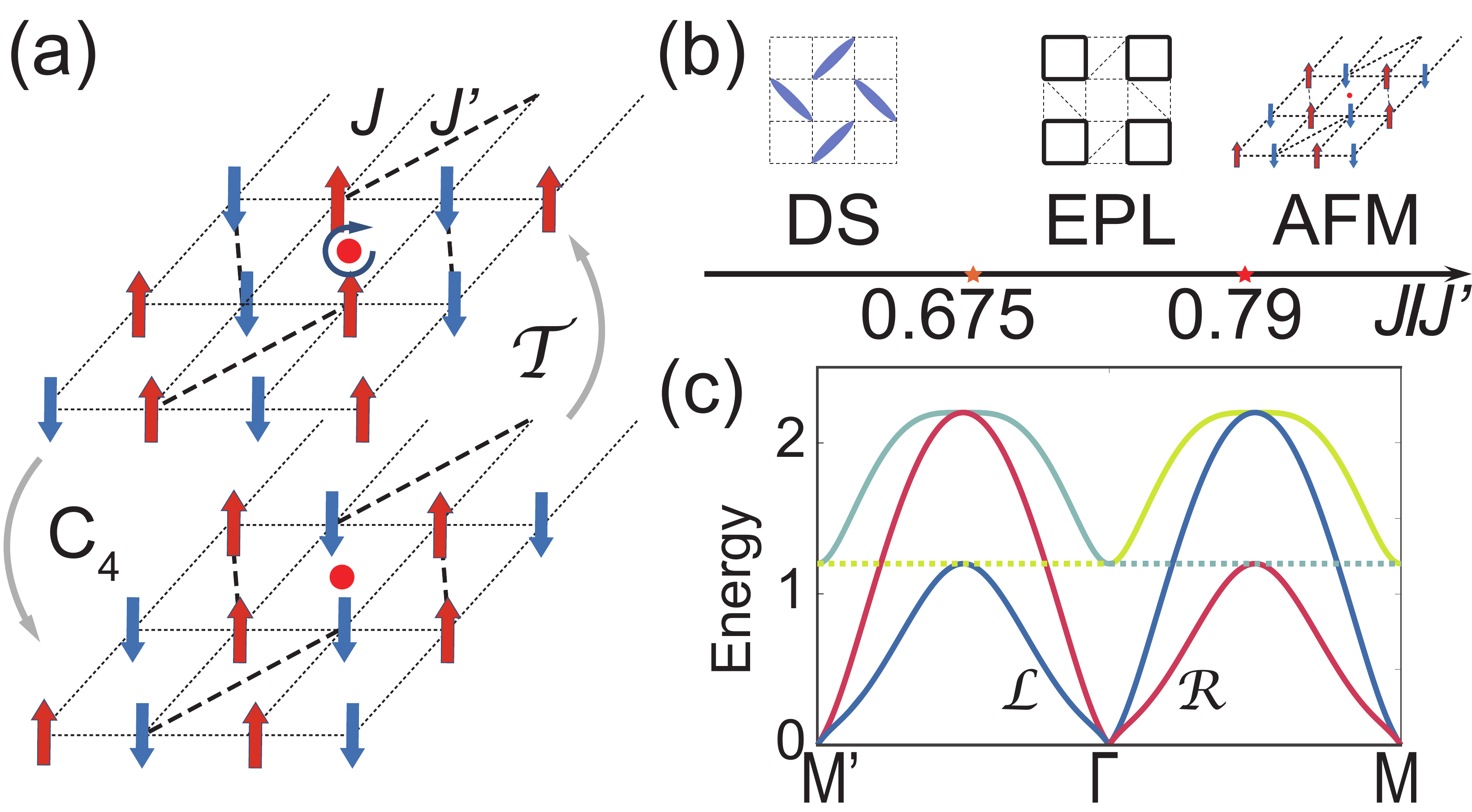}
		\caption{
			(a): Sketch of the SS lattice and the corresponding altermagnetism. The
			two sublattices of the N\'eel state
			are connected by a $C_4$ rotation of the lattice about the center of the
			empty plaquette (without the dimer bond). A time-reversal symmetry then
			recovers
			the N\'eel order. (b): Sketched phase diagram of the SS model with
			$J/J^\prime$.
			The $J/J^\prime$ values at the two transitions are taken from
			Ref.~\cite{Xi2023}.
			(c): LSW dispersion in the AFM state of the SS model showing the chiral
			magnon bands with a non-zero splitting along the M$^\prime$-$\Gamma$-M
			direction of
			the Brillouin zone. $\mathcal{L}$($\mathcal{R}$) refers to the
			left(right)-handed chiral magnon
			bands.}
		\label{fig1}
	\end{figure}

	Tuning the frustration is a promising way to induce a VBS-to-AFM
	transition, and a prominent example is given by the Shastry-Sutherland (SS)
	model~\cite{Shastry1981}. The model is defined on the SS
	lattice as illustrated in Fig.~\ref{fig1}(a).
	It contains orthogonal spin dimers (with coupling $J^\prime$) connected by
	frustrated nearest neighbor (n.n.) bonds (with coupling $J$). With increasing
	$J/J^\prime$, the ground
	state experiences a series of quantum phase transitions from
	a dimer singlet (DS) phase to a plaquette valence bond solid (PVBS) at
	$J/J^{\prime}\approx0.68$, then to a N\'eel AFM state
	for $J/J^{\prime}\sim0.8$ (see Fig.~\ref{fig1}(b))~\cite{Koga2000,
		Chung2001, Pixley2014, Corboz2013, Boos2019, Lee2019, yang2022quantum,
		Sengupta_PRL:2013, Wang_PRL_2023}. Although the DS-to-PVBS
	transition is clearly
	strongly
	first-order, the nature of the PVBS-to-AFM
	transition remains elusive: Several tensor network
	calculations~\cite{Corboz2013,Xi2023} showed it to be weakly first-order, while
	some other \ry{studies~\cite{yang2022quantum, Corboz_arXiv_2025}} found an 
	intervening gapless quantum
	spin
	liquid in between the PVBS and AFM phases. Interestingly, one DMRG
	work~\cite{Lee2019} suggested the transition is via a DQCP with an emergent
	$O(4)$ symmetry, and quantum scaling behavior near this point has been revealed
	by a finite-size tensor network analysis~\cite{WenyuanLiu_PRL_2024}. One
	prominent
	signature of a DQCP is the emergence of deconfined fractionalized spin
	excitations. Although spin excitations in both PVBS and AFM states have
	been
	extensively studied~\cite{Mila_BOT,Pinaki_PRB92_094440_2015, WangZ2018,Wang2023}, whether deconfined excitations emerge
	at the transition is still unclear.
	
	The phase diagram of the SS model well describes the evolution of
	low-temperature
	phases of the quasi-2D antiferromagnet
	$\mathrm{SrCu}_{2}(\mathrm{BO}_{3})_{2}$ under pressure
	tuning~\cite{Kageyama1999,PhysRevLett.82.3701,Koga2000,
		Waki2007,haravifard2016crystallization, Zayed2017, Bettler2020,
		Guo2020,Jimenez2021}, giving the hope to realize a DQCP in this material.
	Remarkably, a recent NMR study in this compound~\cite{Cui2023DQCP} reported an
	unusually large anomalous dimension, $\eta\sim 0.2$, at a field-induced 
	PVBS-to-AFM
	transition under pressure, implying behavior in proximity to a DQCP. However,
	direct spectral evidence on the deconfined excitations at zero field is
	absent.
	
	In this
article, we study spin excitations of the SS model by using
	the state-of-the-art tensor network technique for infinite lattice 
	systems, {\it i.e.}, the infinite projected entangled pair state (iPEPS) 
	method.
	This
	prevents
	the severe finite-size effect in usual numerical calculations and offers a new
	perspective on the nature of emergent DQCP phenomena. \ry{Rather than 
	directly searching for the elusive emergent deconfined excitation continuum 
	at the DQCP, we investigate the evolution of the collective excitation 
	modes (namely, Higgs and Goldstone modes, etc.) associated with the 
	broken 
	emergent enhanced symmetry near the transition in the ordered phases. We 
	demonstrate 
	that their simultaneous, significant softening directly points to 
	the 
	existence of a DQCP.
	In the AFM state, we find
	that the altermagnetic nature of the N\'{e}el order}
	results in a finite splitting with opposite chirality in the magnon bands
	along the $\Gamma$ to M direction of the Brillouin zone. Besides
	the transverse Goldstone excitations, \ry{we 
	have identified two additional low-energy  
	modes: 
	a Higgs mode in the longitudinal excitation channel and an $S=0$ mode with 
	vanishing spectral weight. 
	As the system approaches the PVBS-to-AFM transition, these modes in 
	the 
	AFM phase, 
	as well as the lowest triplet and singlet modes in the plaquette phase, 
	soften and exhibit gap closing behavior at the transition point. We further 
	find the velocities of the singlet, magnon, and Higgs modes become 
	degenerate at the transition point. This, together with the gap closing 
	behavior aforementioned, strongly supports the presence of a DQCP with an 
	emergent O(4) symmetry at the PVBS-to-AFM transition.} 
	
	\section{Model and method}
	The Hamiltonian of the SS model reads
	as
	\begin{equation}\label{Eq:Ham_SS model}
		\mathcal{H} = J \sum_{\langle i,j \rangle} \mathbf{S}_i \cdot \mathbf{S}_j +
		J^\prime \sum_{\langle\langle i,j \rangle\rangle} \mathbf{S}_i \cdot
		\mathbf{S}_j,
	\end{equation}
	where $\mathbf{S}_{i}$ is an $S=1/2$ spin operator defined on site $i$, $J$ and
	$J^\prime$ are
	AFM inter- and intra-dimer couplings on the SS lattice, and we
	set $J^\prime=1$ in
	this work. We apply the 
	iPEPS
	method~\cite{verstraete2004renormalization,orus2014practical,
		corboz2014competing,orus2009simulation,PhysRevB.84.041108} to calculate the
	ground states of the
	model. The iPEPS wavefunction is obtained by variationally minimizing the
	ground state energy using gradient-based optimization, where the gradients are
	calculated through the automatic differentiation technique \cite{SRGAD,
		liao2019differentiable}.
	Then we construct the iPEPS ansatz for excited states through the single-mode
	approximation, and solve the generalized eigenvalue problem for the Hamiltonian
	in the tangent space spanned by these excited states
	~\cite{ponsioen2020excitations,ponsioen2022automatic,chi2022spin}. Details of
	the methods are explained in the
Appendix. With the
	information
	of the excited states, we calculate the spin dynamical structure factor
	(DSF) $\mathcal{S}^{\alpha\beta}(\mathbf{q},\omega) = \frac{1}{N} \sum_{i,j}
	\int dt e^{i\mathbf{q}\cdot(\mathbf{r}_i-\mathbf{r}_j)} e^{i\omega t}
	\langle
	S_i^\alpha (t) S_j^\beta (0) \rangle$, where $\alpha,\beta$ refer to the spin
	components. The excitation spectra are calculated within the tensor network
	states up to
	bond dimension $D=5$. We also compare the results with linear
	spin wave
	(LSW) and bond-operator theories, which are introduced in detail in the 
	Appendix.

	\section{Altermagnetism and chiral magnons in the SS model}
\begin{figure}[t!]
	\centering
	\includegraphics[width=\linewidth]{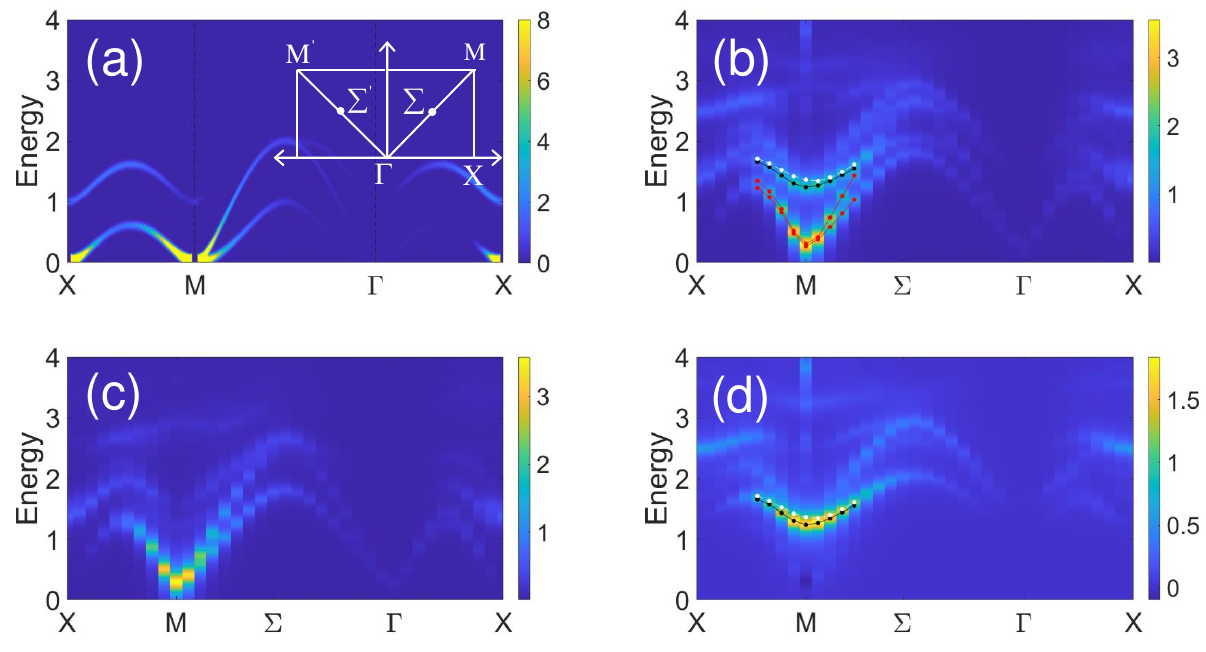}
	\caption{(a): Spin DSF in the AFM phase of the SS model calculated by the
		LSW
		theory. The inset shows high symmetric directions of the Brillouin zone.
		(b): Total
		spin DSF ($\mathcal{S}^{xx+yy+zz}$) in the AFM phase calculated by the 
		iPEPS
		method with $D=3$ at $J/J^\prime=1.1$. (c): The transverse component
		($\mathcal{S}^{xx+yy}$) of the
		results in (b). (d): The longitudinal component ($\mathcal{S}^{zz}$) of 
		the
		results in (b). \ry{
			The red, black, and white lines with dots in (b) and (d) label the 
			magnon, Higgs, and $S=0$ modes, respectively.}
	}
	\label{fig2}
\end{figure}

	The space
	group of the SS lattice is non-symmophic and the two sublattices of the N\'eel
	AFM
	state
	are connected by neither translational nor inversion, but a $C_4$ rotation
	about the center of the empty plaquette. The AFM order then recovers by further
	applying a time-reversal $\mathcal{T}$ symmetry to the spins (see
	Fig.~\ref{fig1}(a)). This indicates that the N\'eel AFM state on the SS lattice
	is an
	altermagnet with a $d$-wave symmetry.
	The spin space
	group symmetry dictates a
	non-zero splitting with opposite chirality of the magnon bands along the
	diagonal directions of the Brillouin zone. This is confirmed by the LSW
	calculation shown in Fig.~\ref{fig1}(c). Along the M$^\prime$-$\Gamma$-M
	direction, the acoustic
	(Goldstone) modes split into two branches with left- and
	right-handed chiralities, respectively. The splitting between the two chiral
	modes along $\Gamma$-M and $\Gamma$-M$^\prime$ directions has opposite sign,
	and its
	magnitude reaches a maximum at the mid-point $\Sigma$ (and $\Sigma^\prime$). On
	the other hand,
	the two modes are degenerate along the $\Gamma$-X (and X-M) directions
	(Fig.~\ref{fig2}(a)). These properties verify that the altermagnet has a
	$d$-wave
	nature. Note that the optical modes also split, but one band is completely flat
	along the M$^\prime$-$\Gamma$-M due to frustration and has zero spectral
	weight.

	\section{Spin excitation spectra}
	We implement
	tensor network calculation
	to go beyond the LSW results on the spin excitation spectra. The calculated
	spectrum at $J/J^\prime=1.1$ is shown in Fig.~\ref{fig2}(b). It exhibits
	richer structures than the LSW results in Fig.~\ref{fig2}(a). To
	understand these structures, we calculate the transverse and longitudinal
	components of the DSF, $\mathcal{S}^{xx+yy}(\mathbf{q},\omega)$ and
	$\mathcal{S}^{zz}(\mathbf{q},\omega)$, and show
	them in Fig.~\ref{fig2}(c) and (d), respectively. The transverse
	DSF resembles the LSW spectrum with three dispersive magnon bands. The
	finite
	gap at M (and $\Gamma$) of the Goldstone modes is an effect of finite
	bond
	dimension $D$ of the iPEPS ansatz, and vanishes in the
	large-$D$ limit (Fig.7 of
the Appendix). One prominent feature is
	the
	splitting of Goldstone modes along the $\Gamma$-M direction. The opposite sign
	of $\Delta\mathcal{S}
	=\mathcal{S}^{+-}(\mathbf{q},\omega)-\mathcal{S}^{-+}(\mathbf{q},\omega)$ for
	the two modes shown in Fig.~8
justifies their opposite chirality, as
	expected for an altermagnet.
	
	The spectrum of longitudinal modes, shown in Fig.~\ref{fig2}(d), also
	consists of multiple bands, owing to the complex SS lattice structure (see
	discussion
	below). We can identify
	a mode (labeled by the black dashed line) that develops a strong 
	resonance at the M
	point. The excitation gap of this mode decreases with reducing $J/J^\prime$ 
	(see Fig.6
	of
the Appendix). We then
	attribute it to the Higgs mode which reflects the amplitude fluctuations of the
	AFM
	order parameter. In a two-dimensional AFM state, the Higgs mode is usually
	hardly visible because the gapless Goldstone modes give rise to a divergent
	longitudinal susceptibility and the Higgs mode can decay into a pair of
	Goldstone modes~\cite{Poldolsky_PRB84_174522_2011}. Here it shows up in the
	tensor network
	calculation
	for the following reasons: The gapped Goldstone modes at finite $D$ remove the
	divergence, and the single-mode approximation adopted suppresses the two-magnon
	continuum but preserves the two-magnon bound state, which is just the Higgs
	mode. \ry{Just slightly above the Higgs mode, we find another  
	excitation mode that is
	quasi-degenerate to the Higgs one at M point but with vanishing spectral 
	weight. The gap of this mode follows a trend similar to that of the Higgs 
	mode as $J/J^\prime$ decreases. We therefore identify this mode as a 
	singlet ($S=0$)
	excitation originating from fluctuations of the plaquette VBS order 
	parameter.} 
	Besides the collective excitation modes discussed above,
	the
	spectrum also contains a continuum, as shown in the broad peak above about
	$2J^\prime$ in the density of states in Fig.~9 of the Appendix.
	Interestingly, the peak moves to lower energy when reducing $J/J^\prime$
	toward the PVBS-to-AFM transition, in a way similar to the behavior of the
	Higgs mode.

	\begin{figure}[t]
		\centering
		\includegraphics[width=\linewidth]{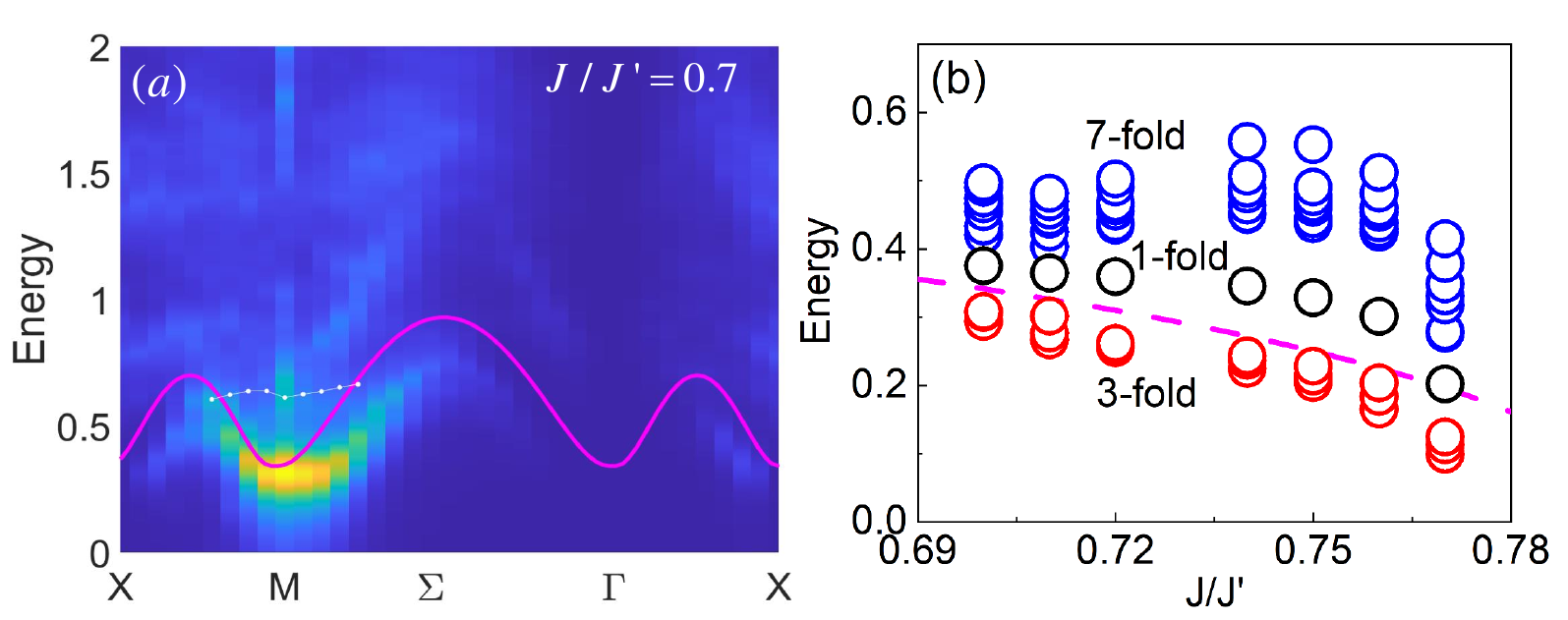}
		\caption{
			(a): Spin DSF in the plaquette phase calculated by the iPEPS method
			with $D=3$
			at $J/J^\prime=0.7$. \ry{The white line with dots shows the lowest 
			singlet excitation. The
			magenta dashed line} labels the dispersion of the triplet 
			excitations in the
			bond-operator theory. (b): $J/J^\prime$ dependence of extracted
			gaps at the M point of the Brillouin zone for several low-energy
			excitation
			modes in the plaquette phase, including, from bottom to top, a 
			triplet (in red),
			a
			singlet (in black), and a 7-fold multiplet (in blue). The results 
			are obtained by the
			iPEPS method with \ry{$D=5$}. The dashed line shows the gap of the
			triplet excitations in the bond-operator
			theory.
		}
		\label{fig3}
	\end{figure}
	
	For a comprehensive understanding, we also study the excitations in the
	plaquette
	phase, and a typical
	spectrum at $J/J^\prime=0.7$ is shown in Fig.~\ref{fig3}(a). The
	lowest-energy
	excitations are found to be a triplet mode, whose dispersion can be
	described by the bond-operator theory~\cite{Mila_BOT, SM}. \ry{Above the 
	triplet mode, we can identify a singlet excitation which is almost 
	dispersionless and has vanishing spectral weight (white dotted line in 
	Fig.~\ref{fig3}(a)).} Fig.~\ref{fig3}(b)
	illustrates
	excitation gaps at the M point of all identified low-energy modes, which, from
	bottom
	to top, include the triplet mode, the singlet mode, and a 7-fold 
	quasi-degenerate multiplet.
	Note that the minor splitting 
    within each triplet
	comes from numerical accuracy and can
	be eliminated by enforcing the spin SU(2) symmetry in the calculation.
	The gap of the triplet mode decreases with increasing
	$J/J^\prime$ toward the PVBS-to-AFM transition and the gap size is
	comparable
	with that from the bond-operator theory.  
	Similar to the triplet, the lowest singlet excitation, which is
		interpreted as PVBS fluctuation, 
		is also softened when approaching the transition. But  
	the 7-fold multiplet does not exhibit significant softening. 
	We calculate the quantum numbers of the 7-fold multiplet and
	find that this multiplet consists of two triplets and one singlet, which are
	the eigenstates of a single plaquette(see Fig.~10 of
the Appendix). The
	inter-plaquette
	interaction then couples
	multiplets in each plaquette into bands that can be understood by a generalized
	bond-operator theory~\cite{Pinaki_PRB92_094440_2015}. Note that each triplet
	excitation
	will further split to transverse and longitudinal modes once the magnetic order
	breaks the spin SU(2) symmetry, accounting for the complex multiband structure
	shown in Fig.~\ref{fig2} in the AFM phase.
	Above these
	low-energy excitations, the spectrum becomes continuous above about
	$J^\prime$ (Fig.~9 of
the Appendix), similar to
	the case of the AFM phase.

	\begin{figure}[t]
	\centering
	\includegraphics[width=0.95\linewidth]{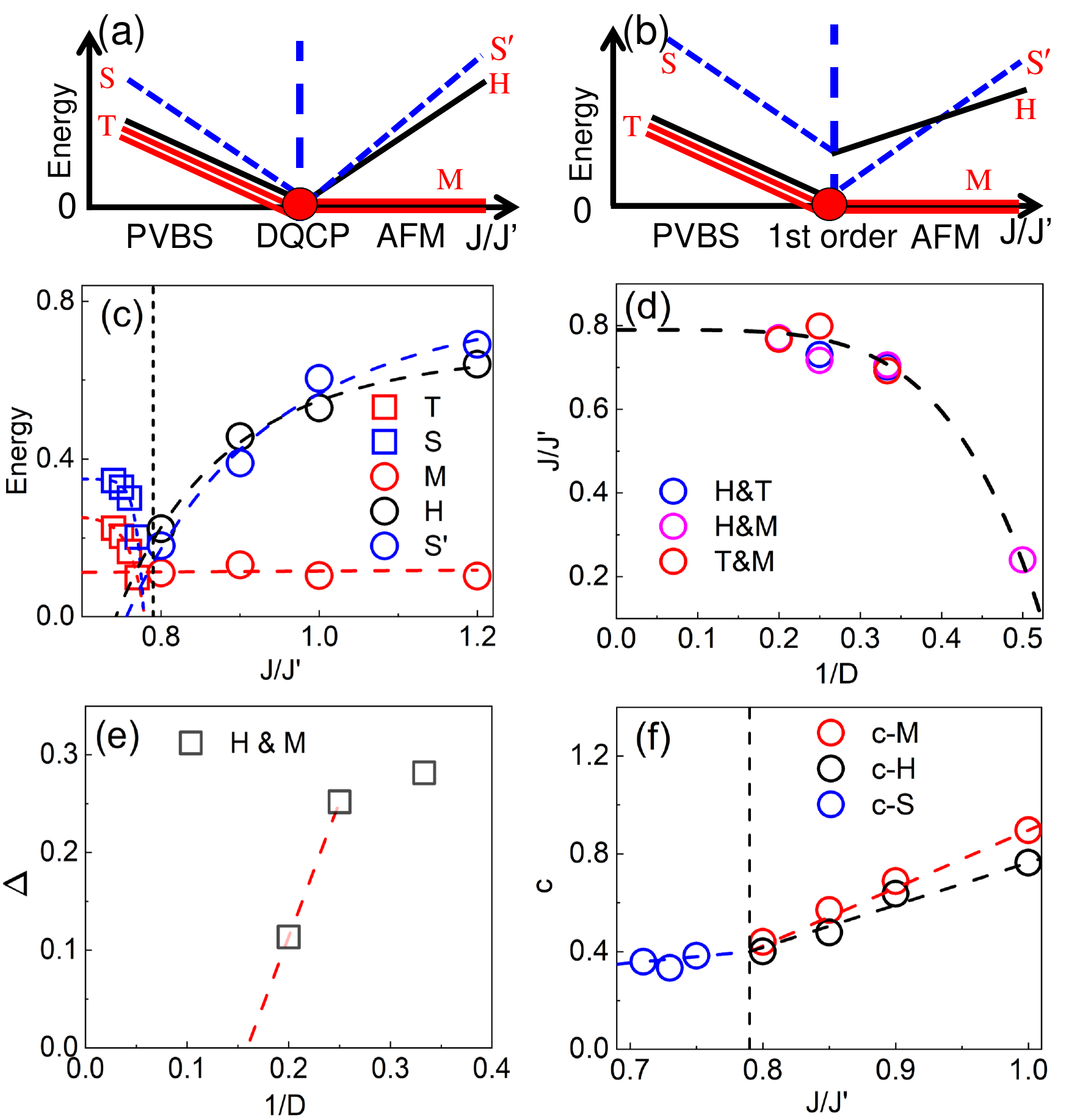}
	\caption{\ry{(a): Evolution of low-energy collective (Higgs and 
			Goldstone) 
			modes across a DQCP with emergent O(4) symmetry. In the PVBS 
			phase, the singlet (S) and triplet (T) modes serve as the Higgs and 
			Goldstone modes associated with the emergent O(4) symmetry. In the 
			AFM 
			phase the Higgs mode (H) turns to a spin longitudinal mode (in 
			black) 
			whereas the singlet (S$^\prime$) and two magnon (M) modes  
			constitute 
			as the Goldstone modes. The DQCP is characterized by simultaneous 
			gap 
			closing of all four modes. (b): Similar to (a), but across a 
			first-order 
			transition with emergent O(4) symmetry. Contrast to the DQCP 
			scenario,  
			the gap of the Higgs mode keeps to be finite across the first-order 
			transition. (c): Gaps at the M point of the triplet (T) and singlet 
			(S) 
			modes 
			in the plaquette phase and of the magnons (M), the Higgs mode (H), 
			and the singlet mode (S$^\prime$) with
			$J/J^\prime$, obtained in the
			iPEPS
			calculation
			with $D=5$. A simultaneous softening of all modes with gaps close 
			to the magnon gap takes place near the PVBS-to-AFM transition 
			at
			$(J/J^\prime)_c\approx0.79$. (d): Finite-$D$ analysis on the 
			crossing points of the
			triplet, Higgs, and magnon modes. In the large-$D$ limit, all
			crossing points are close and can be extrapolated to 
			$(J/J^\prime)_c\approx0.79$, suggesting
			the existence of a DQCP. (e): Finite $D$ analysis of the Higgs gap, 
			$\Delta$, at the crossing
			point with the magnon mode, showing the closure of the Higgs gap in 
			the large $D$ limit.
			(f): Evolution of the velocities $c$ with $J/J^\prime$ for 
			the singlet mode in the PVBS phase together with the Higgs and 
			magnon modes in the AFM phases. $c$ of the these modes 
			coincide at $(J/J^\prime)_c$. For each mode, $c$ is extracted 
			by fitting the dispersion along the M to X direction with 
			$\omega(\mathbf{q})=\sqrt{\Delta^2+c^2q^2}$.}
	}
	\label{fig4}
\end{figure}
	
	\section{Probing the DQCP}
	With excitations studied in
	both PVBS and AFM phases, we now \ry{investigate}
	the PVBS-to-AFM transition. \ry{Though a key characteristic of the 
	DQCP is 
	the emergence of deconfined fractionalized spin excitations, detecting them 
	numerically is, in practice, challenging. This is because they only show up
	at the critical point and manifest as an elusive broad continuum in the 
	spectrum. Here we adopt a different strategy to detect the possible DQCP by 
	investigating the evolution of low-energy collective modes, {\it i.e.} 
	Higgs and Goldstone modes associated with the broken emergent O(4) 
	symmetry, in the PVBS and AFM phases. 
	}  
	
	\ry{
	Let us denote the three components of the 
	AFM order parameter as $n_1$, $n_2$, and $n_3$, and the PVBS order 
	parameter as $n_4$. In the SS model, they are defined as 
	\begin{align}
	 & n_{\alpha=1,2,3} = \sum_{i} (-1)^{i} \langle S_i^{\alpha} \rangle, \\
	 & n_4 = \sum_{i} \left[(-1)^{i_x} \langle \mathbf{S}_i 
	 \mathbf{S}_{i+\hat{x}}\rangle + 
	 (-1)^{i_y} \langle \mathbf{S}_i \mathbf{S}_{i+\hat{y}} \rangle\right]. 	
	\end{align}
	If a DQCP exists, these four components should be combined to a 
	4-dimensional supervector by the emergent O(4) symmetry at the DQCP. When 
	the system is driven to either the PVBS or AFM phase (with either 
	$n_4\neq 0$ or $n_3\neq 0$) under an infinitesimal perturbation, the 
	emergent O(4) symmetry is broken 
	to O(3). This results in three Goldstone modes and one Higgs mode in 
	each phase. In the AFM phase the Higgs mode is in the longitudinal spin 
	($S^z$) channel, whereas in the PVBS phase it corresponds to the 
	amplitude fluctuations of $n_4$ which is a spin singlet. The Goldstone 
	modes are created by the generators of the O(4) algebra that rotate the 
	ordered component to 
	other three disordered components. In the PVBS phase, $n_4$ is ordered, 
	and $n_1$, $n_2$, and $n_3$ follow the fundamental spin SO(3) symmetry. As 
	a result, the three Goldstone modes are just the lowest spin triplet 
	excitations. Note that the triplet excitations are generally gapped since 
	the PVBS order breaks discrete lattice symmetry, and strictly 
	speaking, they are pseudo-Goldstone modes (in the context of emergent O(4) 
	symmetry). But as the system 
	approaches the critical 
	point, they should exhibit a gap closing behavior and develop as the 
	emergent gapless Goldstone modes. In the AFM phase, $n_3$ is ordered and in 
	the 
	vicinity of the critical point, $n_1$, $n_2$, and $n_4$ should follow an 
	emergent O(3) symmetry. Besides the two magnons, which are Goldstone modes 
	associated with $n_1$ and $n_2$, respectively, there is an additional 
	emergent 
	Goldstone mode 
	associated with $n_4$, which is a spin singlet with a gap 
	closure at the DQCP.}
	    
	\ry{The evolution of these collective modes across an O(4) DQCP is sketched 
	in Fig.~\ref{fig4}(a). Approaching the critical point, three Goldstone 
	modes and a Higgs mode soften, and their energies all go to zero at the 
	DQCP. Given the O(4) symmetry at the DQCP, their difference disappears 
	and they turn to 4-fold degenerate critical modes forming the lower edge of 
	the continuum of deconfined spinons. This scenario contrasts to that of a 
	first-order transition with an emergent O(4) symmetry. In the 
	latter case, the composite 4-component order parameter is non-zero 
	at the transition point. This breaks the O(4) symmetry in the ground state, 
	giving rise to 
	three emergent Goldstone modes and a gapped Higgs mode as 
	illustrated in Fig.~\ref{fig4}(b).  
}

\ry{To examine which of the above two scenarios applies to the SS model, we 
have extracted gaps 
of the low-energy collective excitations at the M point, and their evolution 
across the 
PVBS-to-AFM transition for $D=5$ is plotted in  
	Fig.~\ref{fig4}(c). 
The gaps for other $D$ values are presented in Fig.~11 of
	the
Appendix. 
The magnon gap is almost independent of $J/J^\prime$, confirming that it is
	caused by finite $D$, which closes in the entire AFM phase in the large $D$ 
	limit (see 
	Fig.~7 of the 
	Appendix). On the other hand, the gaps of all other excitations, including 
	the Higgs and singlet modes in the AFM phase and the triplet and singlet 
	modes in the plaquette phase, drop rapidly and are close to the magnon gap 
	when approaching the transition point 
	$(J/J^\prime)_c\approx0.79\pm0.02$.  
	As shown in Fig.~\ref{fig4}(d), 
	the crossing points of these modes are all close
	and can be extrapolated to $(J/J^\prime)_c$ in the large-$D$ limit. This 
	implies that there is a single transition at $(J/J^\prime)_c$ with gap 
	closure of all excitations. We then plot
	the gap of the Higgs mode at the crossing point with 
	magnons in Fig.~\ref{fig4}(e). The Higgs gap indeed closes in 
	the large $D$ limit, corroborating the DQCP scenario.
}

\ry{In Fig.~\ref{fig4}(f) we further plot the extracted velocities $c$ of 
several 
excitation modes by fitting their dispersions along the M to X direction with 
$\omega(\mathbf{q})=\sqrt{\Delta^2+c^2q^2}$. Besides the gap closure behavior, 
we find the velocities of the singlet mode in the plaquette phase and the 
magnon and Higgs modes in the AFM phase coincide at $(J/J^\prime)_c$. This once 
again supports the transition is via a DQCP: At the critical point, both
Higgs and Goldstone modes become degenerate and appear as the lower edge 
of the deconfined spinon continuum $\omega_{\text{min}}(\mathbf{q})\sim c 
|\mathbf{q}|$, where the velocity $c$ is universal for all critical modes due 
to the emergent Lorentz invariance. The degeneracy of 4 modes -- 1 from the 
$S=0$ 
(singlet) and 3 from the $S=1$ (Higgs and magnons) channel -- supports the 
emergent symmetry of the DQCP is O(4).	
}
		
	\section{Discussions and conclusions}
    \ry{The appearance of 4 degenerate 
	low-lying excitation modes in both $S=0$ and $S=1$ channels when 
	approaching the transition from either side 
	as well as the gap closure of the Higgs mode provide direct evidence 
	that the transition goes beyond the conventional O(3) Wilson-Fisher 
	universality or a symmetry-enhanced first-order transition, but is instead 
	via a DQCP with emergent O(4) symmetry. Quantum scaling behavior with novel 
	critical exponents characterizing this DQCP is then 
	expected~\cite{WenyuanLiu_PRL_2024}. Some recent 
	studies~\cite{yang2022quantum, 
	Corboz_arXiv_2025} suggest that a quantum spin liquid can be stabilized in 
	between the PVBS and AFM phases. Our analysis on the spin 
	excitations cannot rule out a spin liquid phase, 
	yet implies a spin liquid could only stabilized within a narrow parameter 
	regime $0.77\lesssim J/J^\prime \lesssim 0.81$. 
	Our calculation suggests the 4 critical modes keeps to be the only 
	low-lying excitations within this parameter regime. Such a few number of 
	low-lying excitations rules out the possibility of U(1) spin 
	liquids~\cite{Song_NC_2019}, and 
	the putative spin liquid is most likely a $Z_2$ Dirac spin 
	liquid~\cite{Maity_arXiv_2025}, 
	in which 
	the $S=0$ excitation turns out to be the vison and opens a small gap while 
	the other three modes represent gapless excitations in the spinon 
	sector.   
}
	
	\ry{Moreover, our scenario naturally applies to a large class of models
	including the SS model, the checkerboard $J$-$Q$ model,
	and their variants.
	The PVBS-to-AFM transition in the checkerboard $J$-$Q$ model is found 
	to be first-order with an enhanced O(4) symmetry. The low-lying 
	excitations can then be described by the proximate DQCP scenario shown in 
	Fig.~\ref{fig4}(b). While the Higgs mode is gapped at the transition, 
	the emergent Goldstone modes cause a gap closure and are responsible  
	for the observed
	quantum scaling behavior~\cite{Zhao_NP_2018,ChengchenLi_JPCM_2024}. Recent 
	studies proposed a 
	global 
	phase diagram where the DQCP plays as a fine tuned multicritical point 
	connecting the 
	PVBS, 
	AFM, and quantum spin liquid phases~\cite{Zhao_PRL_2020, Lu_PRB_2021}. Our 
	calculation suggests that the PVBS-to-AFM transitions in the SS model 
	is either at or very close to the multicritical point whereas the one in 
	the checkerboard $J$-$Q$ model is at a different regime of this proposed 
	phase diagram, via a symmetry enhanced weakly first-order transition. 
	Note that some other studies suggest the underlying conformal field theory 
	(CFT) for the SO(5) DQCP could be 
	non-unitary~\cite{ma2020theory,nahum2020note,zhou2024so}.  
	Given that the O(4) DQCP is obtained by deforming the SO(5) 
	CFT~\cite{Lee2019}, this implies that realization of an O(4) DQCP on a 
	lattice model without fine tuning could be 
	delicate, which is reconciled with the proposed multicriticality picture.
}

\ry{In either the SS or the checkerboard $J$-$Q$ model, $C_4\mathcal{T}$ 
symmetry	
dictates the N\'{e}el AFM state of the system to be a $d$-wave altermagnet 
with split magnon bands. One might worry that the splitting of magnon bands 
could break the emergent Lorentz invariance and thereby turn the PVBS-to-AFM 
transition into a weakly first-order one. However, the magnon dispersion reads 
$\omega(\mathbf{q})=cq \pm w q_x q_y$, where the splitting 
is caused by the anisotropic $w$ term in the $q^2$ order. 
The magnon bands are still degenerate up to the leading linear-in-$q$ term. 
Therefore, the splitting, though a symmetry-breaking perturbation, is 
irrelevant in the low-energy effective theory of DQCP. Nonetheless, it provides 
a useful probe for the PVBS-to-AFM transition by signaling 
the symmetry breaking characteristic of an altermagnet.}

\ry{We then discuss the implications of our results for experiments on the SS 
material SrCu$_2$(BO$_3$)$_2$. A recent specific heat measurement find a weakly 
first-order PVBS-to-AFM transition at zero field~\cite{LilingSun_arXiv_2024}, 
and the 
quantum scaling of 
the NMR spin-lattice relaxation rate in a magnetic field suggests the 
transition is proximate to a DQCP~\cite{Cui2023DQCP}. It is likely 
that the subleading interactions in the material, which goes beyond the 
description of 
the SS model, such as the Dzyaloshinsky-Moriya interaction or the interlayer 
coupling, drive the transition to first-order but still proximate to a 
putative DQCP. This well explains the unusually large anomalous dimension 
$\eta$ observed in the NMR measurement~\cite{Cui2023DQCP}.}
	
	Besides the possible DQCP, another unsettled
	issue
	is the nature of the PVBS state in experiment. Recent inelastic neutron 
	scattering
	(INS)~\cite{Zayed2017} and
	NMR~\cite{Cui2023DQCP} measurements suggest a full plaquette phase where the
	spin singlets
	are located on plaquettes
	with $J^\prime$ dimers. But most numerical studies (including ours) on the 
	SS
	model
	found an empty plaquette (singlets on plaquettes without the $J^\prime$ 
	dimers) ground state.
	Interestingly, for the plaquette state at $J/J^\prime=0.7$, if taking
	$J^\prime\sim 4$
	meV estimated from the experiment, the calculated spin gaps of the
	lowest
	triplet mode and the 7-fold degenerate mode (including two degenerate triplets)
	are about
	1 meV and 2 meV, which are consistent with the gap values of the lowest two
	excitations observed in the INS experiment~\cite{Zayed2017}. It would be
	important
	to further compare the theoretical and experimental results on the dispersions
	and evolution of the spin gaps with pressure.
	
	We have examined the lattice structure of the SS model
material SrCu$_2$(BO$_3$)$_2$, and find that it indeed supports an
altermagnet even when the non-magnetic ions are included.
	To experimentally probe the altermagnetism, one may look for the
	splitting of the two magnon bands (at the order of $J^\prime$ near the
	$\Sigma$ point) in the spectrum of INS. The chiral magnons can be detected by
	resonant
	inelastic X-ray scattering (RIXS). Moreover, the transverse spin current
	induced by the chiral magnons will cause sizable spin Nernst effect in the AFM
	state~\cite{Smejkal_PRX_2022B,Cui_PRB_2023}
	when applying a temperature gradient along the non-splitting direction. Given
	that the Dzyaloshinsky-Moriya interaction is much smaller than $J$ and
	$J^\prime$ in SrCu$_2$(BO$_3$)$_2$, the leading contribution should be
	attributed to the altermagnetism. Given that the splitting of the magnon
	bands
	in the altermagnetic phase is directly related to the O(4)
	symmetry breaking in the magnetic sector, the proposed experiments above,
	provide viable
	ways to detect the proximate deconfined quantum criticality.
	
	\ry{In conclusion, we study the spin excitations of the SS model by
	using advanced tensor network method and examine the evolution of 
	low-energy collective modes across the transition between the plaquette 
	valence bond solid and the N\'{e}el 
	antiferromagnetic phase. We show that the 
	antiferromagnetic state
	of the model is an altermagnet supporting two split magnons (Goldstone modes)
	with opposite chirality.
	In addition, we have also identified a Higgs
	mode in the longitudinal spin channel and a spin singlet mode with 
	vanishing spectral weight. The low-energy excitations in the plaquette 
	valence 
	bond solid
	contain 
	a triplet mode 
	and a singlet one.
    As the system approaches the transition, all low-energy collective 
    excitations soften and their gaps close. Simultaneously, their velocities 
    coalesce. These observations demonstrate that the transition is governed by 
    a universal deconfined quantum criticality with an emergent O(4) symmetry.
	Our results clarify the spectral signatures in the vicinity of a deconfined 
	quantum critical point and provide viable means of detecting this exotic 
	phase transition.}
	
	\begin{acknowledgments}
	We thank Runze Chi, Wenan Guo, Bruce Normand, Wei Li, Zhengxin Liu, Anders W.
	Sandvik,
	Yiming Wang, Jize Zhao for helpful discussions.
	This work is supported by the National Key R\&D Program of China (Grant No. 2023YFA1406500),
	the National Natural Science Foundation of China (Grant Nos.~12334008,
	12274458, 12174441).
\end{acknowledgments}



	\appendix

	\section{iPEPS method for calculating the ground state and excited states}
	\begin{figure}[h]
		\centering
		\includegraphics[width=1\linewidth]{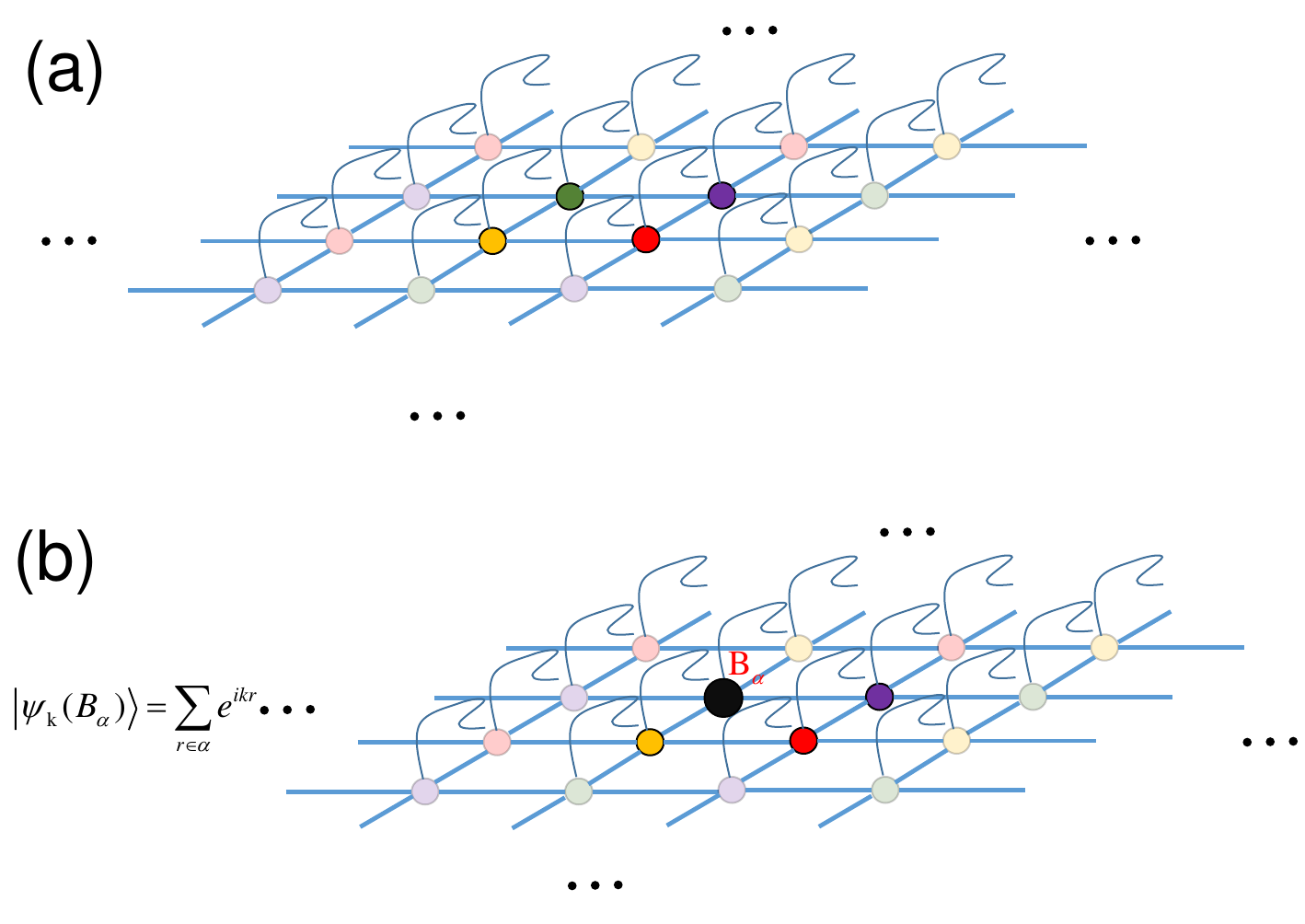}
		\caption{(a): The iPEPS setup in the calculation of the ground state
			of the SS model. The tensors are arranged at the sites of the SS lattice. The bright-colored sites indicate a $2\times2$ unit cell used in the calculation. (b):
			Illustration of the single-mode iPEPS ansatz for the calculation of excited
			states. In each term of the summation, a single local tensor $A_\alpha$ located at site $r$ belonging to a specified sublattice is modified to a different tensor $B_{\alpha}$. Here $\alpha=1,2,3,4$, corresponding to a $2\times 2$ unit cell.}\label{fig:S1}
	\end{figure}	
	
	In this work, we apply the infinite projected entangled pair state (iPEPS) method~\cite{verstraete2004renormalization,orus2014practical} to calculate the ground state of the Shastry-Sutherland (SS) model. A sketched iPEPS setup with a $2\times2$ unit cell is shown in
	Fig.~\ref{fig:S1}(a). In the iPEPS setup, a local tensor is defined on each site of
	the SS lattice. As shown in Fig.~~\ref{fig:S1}, each local tensor has a physical index corresponding to the spin degrees of freedom, with dimension $2$ and denoted as curve, and four virtual indices corresponding to the four links of the site, with dimension $D$ referred to as the iPEPS bond dimension and denoted as straight. In this work, as emphasized as bright color in Fig.~\ref{fig:S1}, the unit cell is chosen to be $2\times 2$, and this is the minimal size compatible with the ground state of the SS model.
	
	Given an iPEPS state $|\Psi\rangle$, the ground state energy can be evaluated according to the fundamental quantum mechanics
	\begin{eqnarray}
		E &=& \frac{\sum_{\langle ij\rangle}\langle\Psi|\hat{H}_{ij}(J)|\Psi\rangle}{\langle\Psi|\Psi\rangle}\nonumber\\
		 &+& \frac{\sum_{\langle\langle ij\rangle\rangle}\langle\Psi|\hat{H}_{ij}(J^\prime)|\Psi\rangle}{\langle\Psi|\Psi\rangle} \label{Energy}
	\end{eqnarray}
	where $\langle ij\rangle$ and $\langle\langle ij\rangle\rangle$ denote nearest neighbors and second nearest neighbors, respectively. In this sense, the energy can be regarded as a complicated function of the local tensors, the variational parameters in the iPEPS ansatz. The denominators and numerators in Eq.~(\ref{Energy}) can be effectively calculated by using the corner transfer matrix renormalization group method \cite{corboz2014competing,orus2009simulation,PhysRevB.84.041108, VCTMRG2022}. Then, starting from an arbitrary state, the ground state is obtained by variationally minimizing the energy through gradient-based optimization, e.g., the L-BFGS optimization strategy \cite{LBFGS}. In this work, the energy gradients concerning the local tensors are obtained effectively by the automatic differentiation techniques \cite{SRGAD, liao2019differentiable}.
	
	Once the ground state $|\Psi_0\rangle$ is obtained, we can construct the iPEPS ansatz for the excited state through the single-mode approximation \cite{ponsioen2020excitations,ponsioen2022automatic,chi2022spin}. More specifically, the excited ansatz with lattice momentum $k$ reads as
	\begin{eqnarray}
		|\Psi_{k}(B_{\alpha})\rangle = \sum_{r\in \alpha}e^{ikr}|\Psi_{r}(B_{\alpha})\rangle \label{ansatz}
	\end{eqnarray}
	where $\alpha$ denotes the four sublattices corresponding to the $2\times 2$ unit cell, and the summation is over all the sites belonging to the $\alpha$-th sublattice. Here $|\Psi_{r}(B_{\alpha})\rangle$ is the perturbed $|\Psi_0\rangle$ state obtained by replacing the local tensor $A_{\alpha}$ at $r$ in $|\Psi_0\rangle$by a different tensor $B_{\alpha}$, which satisfies the orthogonal condition
	\begin{eqnarray}
		\langle\Psi_r(B_{\alpha})|\Psi_0\rangle = 0. \label{Tangent}
	\end{eqnarray}
	In practice, $B_{\alpha}$ can be chosen to be a vector in the null space of $E_{\alpha}$, which is the effective environment of $A_{\alpha}$ in $\langle\Psi_0|\Psi_0\rangle$ and satisfies $\langle\Psi_0|\Psi_0\rangle = \mathrm{Tr}E_{\alpha}A_{\alpha}$. For a given ground state with bond dimension $D$, we have $2D^4-1$ choices of $B_{\alpha}$ satisfying Eq.(\ref{Tangent}) for each $\alpha$, and thus $8D^4-4$ basis vectors, denoted as $|\Phi_{i}(k)\rangle$, in total to span the tangent space of the ground state. Then, we can calculate the matrix representation of the Hamiltonian in this non-orthogonal basis, and solving the generalized eigenvalue equation
	\begin{eqnarray}
		\langle \Psi_{i}(k)|\hat{H}|\Psi_{j}(k)\rangle = E(k)\langle \Psi_{i}(k)|\Psi_{j}(k)\rangle \label{Geig}
	\end{eqnarray}
	gives the excitation energies $E(k)$s and excited eigenstates $|\Phi_i(k)\rangle$s at momentum $k$ . In this work, we have employed an effective derivative trick and a regularization procedure, introduced in Ref.~\cite{ponsioen2020excitations, ponsioen2022automatic}, to determine the matrices that appear in Eq.~(\ref{Geig}).
	
	As long as the wave functions of the excited states $|\Phi_i(k)\rangle$s are obtained, we can use them as a complete set to calculate the spin dynamical structure factor (DSF) ($\alpha = x, y, z$)
	\begin{eqnarray}
		\mathcal{S}^{\alpha\alpha}(k,\omega) &=& \int dt e^{-i\omega t}\langle\Psi_0|S_{-k}^{\alpha}S^{\alpha}_k(t)|\Psi_0\rangle \nonumber \\
		&=& \sum_{mn}\int dt e^{-i(\omega+E_0)t}\langle\Psi_0|S_{-k}^{\alpha}|\Phi_m(k)\rangle \nonumber\\
		&\cdot&\langle\Phi_m(k)|e^{iHt}|\Phi_n(k)\rangle\langle\Phi_n(k)|S^{\alpha}_k|\Psi_0\rangle \nonumber \\
		&=& \sum_{n}\delta(E_n-E_0-\omega)w^\alpha_n(k)
	\end{eqnarray}
	where $w^\alpha_n(k)\equiv|\langle\Phi_n(k)|S_k^\alpha|\Psi_0\rangle|^2$ is the spectral weight for each mode. If regarded as the overlap of two excited-state wave functions, the spectral weight can be tackled similarly by the derivative trick \cite{ponsioen2022automatic}. In practical calculations, a Lorentzian broadening is employed to deal with the delta function and mimic the finite temperature effect.

	\section{Linear spin wave and bond-operator theories for the SS model}
	
	To better understand the tensor network results in the antiferromagnetic (AFM) and the
	plaquette phases, we study the spin excitations in these two phases by
	using the linear spin wave (LSW) and bond-operator theories, respectively.
	
	\subsection{Linear spin wave theory for spin excitations in the
		antiferromagnetic phase}
	
	Here we provide the LSW theory for spin
	excitations in the AFM phase of the SS model.
	We pick up a $2\times2$ magnetic unit cell in the AFM state. Each
	spin on the SSL can be identified by the combination of a vector
	${\mathbf{r}}$ labeling the unit cell position and a sublattice index $l$
	($l=1,2,3,4$) within each unit cell. We introduce a unit vector
	$\mathbf{n}_{l}$ to denote the orientation of the spin at sublattice $l$, and
	then define two additional unit vectors $\mathbf{u}_{l}$ and $\mathbf{v}_{l}$
	according to ${\mathbf{u}_{l}\cdot\mathbf{n}_{l}=0}$
	and ${\mathbf{v}_{l}=\mathbf{n}_{l}\times\mathbf{u}_{l}}$, namely,
	$\mathbf{n}_{l}$, $\mathbf{u}_{l}$ and $\mathbf{v}_{l}$
	are orthogonal to each other. Next we perform the Holstein-Primakoff
	transformation for the spin operator $\mathbf{S}_{{\mathbf{r}}l}$,
	\begin{eqnarray}
		\mathbf{n}_{l}\cdot\mathbf{S}_{{\mathbf{r}}l} & = &
		S-b_{{\mathbf{r}}l}^{\dagger}b_{{\mathbf{r}}l},\\
		(\mathbf{u}_{l}+i\mathbf{v}_{l})\cdot\mathbf{S}_{{\mathbf{r}}l}
		& = &
		({2S-b_{{\mathbf{r}}l}^{\dagger}b_{{\mathbf{r}}l}})^{\frac{1}{2}}b_{{\mathbf{r}}l},\\
		(\mathbf{u}_{l}-i\mathbf{v}_{l})\cdot\mathbf{S}_{{\mathbf{r}}l}
		& = &
		b_{{\mathbf{r}}l}^{\dagger}({2S-b_{{\mathbf{r}}l}^{\dagger}b_{{\mathbf{r}}l}})^{\frac{1}{2}}.
	\end{eqnarray}
	
	Due to the bipartite lattice structure, we adopt
	two sets of Fourier transforms. On the sublattice $l$ with an up spin:
	
	\begin{equation}
		b_{{\mathbf{r}}l} =\sqrt{\frac{2}{N}}\sum_{\mathbf{k}\in{\text{MBZ}}}
		b_{\mathbf{k}l}e^{i\mathbf{R}_{{\mathbf{r}}l}\cdot\mathbf{k}}
	\end{equation}
	and on the sublattice $l^\prime$ with a down spin:
	\begin{equation}
		b_{{\mathbf{r}}l^\prime} =\sqrt{\frac{2}{N}}\sum_{\mathbf{k}\in{\text{MBZ}}}
		b_{\mathbf{k}l^\prime}e^{-i\mathbf{R}_{{\mathbf{r}}l^\prime}\cdot\mathbf{k}}
	\end{equation}
	where $\mathbf{R}_{{\mathbf{r}}l}$ denotes the position of a spin
	on the SSL labeled by the magnetic unit cell position ${\mathbf{r}}$ and
	sublattice index $l$. The spin Hamiltonian can be rewritten in terms of
	boson bilinears as
	\begin{equation}
		H_{\text{LSW}} =\sum_{\mathbf{k}\in{\text{MBZ}}}
		\Psi(\mathbf{k}){}^{\dagger}h(\mathbf{k})\Psi(\mathbf{k})+const.,
	\end{equation}
	where
	\begin{equation}
		\Psi(\mathbf{k})
		=[b_{\mathbf{k}1},b_{\mathbf{k}2},b^\dagger_{\mathbf{k}3},b^\dagger_{\mathbf{\mathbf{k}}4}]^{T},
	\end{equation}
	and $h(k)=$
\begin{equation}
		\left(\begin{array}{cccc}
			2J-\frac{J^\prime}{2} & \frac{J^\prime}{2}e^{\frac{i(k_{x}-k_{y})}{2}} &
			J\cos(\frac{k_{x}}{2}) & J\cos(\frac{k_{y}}{2})\\
			\frac{J^\prime}{2}e^{\frac{-i(k_{x}-k_{y})}{2}} & 2J-\frac{J^\prime}{2} &
			J\cos(\frac{k_{y}}{2}) & J\cos(\frac{k_{x}}{2})\\
			J\cos(\frac{k_{x}}{2}) & J\cos(\frac{k_{y}}{2}) &
			2J-\frac{J^\prime}{2} & \frac{J^\prime}{2}e^{\frac{-i(k_{x}+k_{y})}{2}}\\
			J\cos(\frac{k_{y}}{2}) & J\cos(\frac{k_{x}}{2}) &
			\frac{J^\prime}{2}e^{\frac{i(k_{x}+k_{y})}{2}} & 2J-\frac{J^\prime}{2}
		\end{array}\right).
	\end{equation}
	
	Then we can diagonalize $H_{\text{LSW}}$ via the Bogoliubov transformation with
	${\Psi(\mathbf{k})=T_{\mathbf{k}}\Phi(\mathbf{k})}$,
	where
	\begin{equation}
		\Phi(\mathbf{k}) =[\beta_{\mathbf{k}1},
		\beta_{\mathbf{k}2},\beta_{\mathbf{k}3}^\dagger,\beta_{\mathbf{k}4}^\dagger]^{T},
	\end{equation}
	\color{black}
	is the diagonalized basis and $T_{\mathbf{k}}$ is the transformation
	matrix.
	
	The general form of spin-wave dispersion is complex but can be easily computed
	numerically. Here we present the analytical form of the dispersions of the two
	Goldstone modes along the
	$\Gamma$-M direction near the $\Gamma$ (and equivalently, M) point. Up to the
	second-order terms in the wave vector $k_{x}=k_{y}$, the dispersions for
	$J\neq J^\prime$ are
	\begin{align}
		\begin{matrix}
			E_{+} &
			=\sqrt{\frac{2J^{2}(J-J^\prime)}{2J-J^\prime}}k_{x}
			+\frac{J^{2}J^\prime}{2(2J-J^\prime)^{2}}k_{x}^{2},\\ \\
			E_{-} &
			=\sqrt{\frac{2J^{2}(J-J^\prime)}{2J-J^\prime}}k_{x}
			-\frac{J^{2}J^\prime}{2(2J-J^\prime)^{2}}k_{x}^{2},
		\end{matrix}
	\end{align}
	respectively. Note that the coefficients of the linear-in-$k_x$ terms are the
	same, while the splitting caused by the altermagnetism appears in the
	subleading $k_x^2$ terms.
	
	At $J=J^\prime$ where the AFM ground state becomes unstable in the LSW theory,
	these two magnon bands are softened with dispersion relations
	\begin{align}
		\begin{matrix}
			E_{+} & =\frac{Jk_{x}^{2}}{2},\\ \\
			E_{-} & =0.
		\end{matrix}
	\end{align}
	.
	
	\subsection{Bond operator theory for the triplet excitations in the plaquette
		phase}
	
	In the plaquette phase, the spin rotational symmetry preserves and the LSW theory is
	not applicable. To derive the low-lying triplet excitations in this phase, we
	project the Hamiltonian onto the low-energy subspace composed of
	the singlet ground state and the lowest triplet excitations.
	
	Define a vacuum
	$|0\rangle$ state
	and four boson operators which yield the four physical states (one singlet and
	three triplets) by
	$|s\rangle=s^{\dagger}|0\rangle$,
	$|t_{\alpha}\rangle=t_{\alpha}^{\dagger}|0\rangle$, respectively \cite{sachdev1990bond,zhitomirsky1996valence}. The projection
	operators are expressed $Z^{st_{\alpha}}=s^{\dagger}t_{\alpha}$,
	$Z^{t_{\alpha}t_{\beta}}=t_{\alpha}^{\dagger}t_{\beta}$. The spins
	represented via these boson operators are
	\begin{equation}
		S_{l}^{\alpha}=\frac{(-1)^{l}}{\sqrt{6}}(s^{\dagger}t_{\alpha}
		+t_{\alpha}^{\dagger}s)-\frac{i}{4}\sum_{\beta,\gamma}
		\epsilon_{\alpha\beta\gamma}t_{\beta}^{\dagger}t_{\gamma}
	\end{equation}
	where $\alpha,\beta,\gamma=x,y,z$ denote the spin components and $l$ is the
	sublattice index $(l=1,2,3,4)$.
	Note that the Hilbert space of the bosons is larger than the original plaquette
	Hilbert space and includes unphysical states. To restrict the boson Hilbert
	space to its physical sector, a hard-core constraint must be imposed:
	$s^{\dagger}s+\sum_{\alpha}t_{\alpha}^{\dagger}t_{\alpha}=1$.
	By condensing the singlets $s^{\dagger}=s\approx\langle s\rangle$, we rewrite
	the Hamiltonian in terms of triplet operators
	\begin{equation}
		H=H_{0}+\sum_{\mathbf{k}\in{\text{MBZ}}}\Psi(\mathbf{k}){}^{\dagger}
		h(\mathbf{k})\Psi(\mathbf{k})+const.,
	\end{equation}
	where
	$H_{0}=E_{s}s^{\dagger}s+\sum_{\alpha}E_{\alpha}t_{\alpha}^{\dagger}t_{\alpha}$
	is the energy of a single plaquette and
	\begin{equation}
		\Psi(\mathbf{k})=[t_{\mathbf{k}x},t_{\mathbf{k}y},t_{\mathbf{\mathbf{k}}z},t^\dagger_{\mathbf{-k}x},t^\dagger_{\mathbf{-k}y},t^\dagger_{\mathbf{\mathbf{-k}}z}]^{T}
	\end{equation}
	Then following the standard Bogoliubov diagonalization we can derive
	the dispersion of the lowest triplet as
	\begin{equation}
		E=\sqrt{J^{2}+\frac{2}{3}J(J^\prime-2J)(\cos(k_{x})+\cos(k_{y}))}.
	\end{equation}
	We compare the triplet dispersion with that calculated by using the tensor network method
	in Fig.~3 of the main text, and they agree well. The small difference is likely
	caused by the finite bond dimension $D$ in the tensor network calculation.
	
	\section{Numerical results on the spin excitation spectra}
	
	\begin{figure}[!h]
		\centering
		\includegraphics[width=1\linewidth]{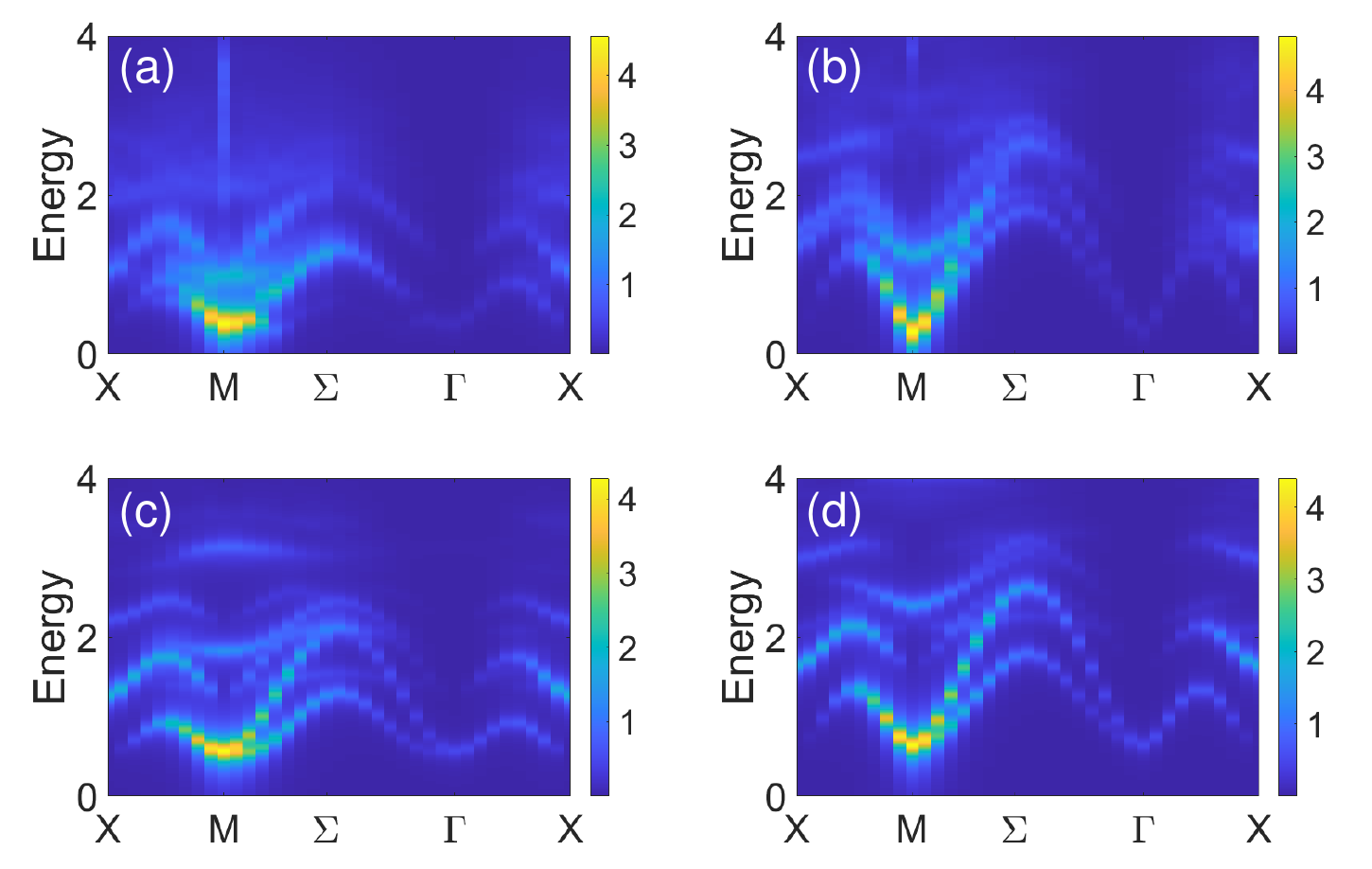}
		\caption{Total
			spin DSF ($\mathcal{S}^{xx+yy+zz}$) in the AFM phase of the SS model calculated by iPEPS
			method with $D=3$ (in (a) and (b)) and $D=2$ (in (c) and (d)), at $J/J^\prime=0.9$ (in (a) and (c)) and $J/J^\prime=1.1$ (in (b) and (d)).}\label{fig:S2}
	\end{figure}
	
	\begin{figure}[!h]
		\centering
		\includegraphics[width=1\linewidth]{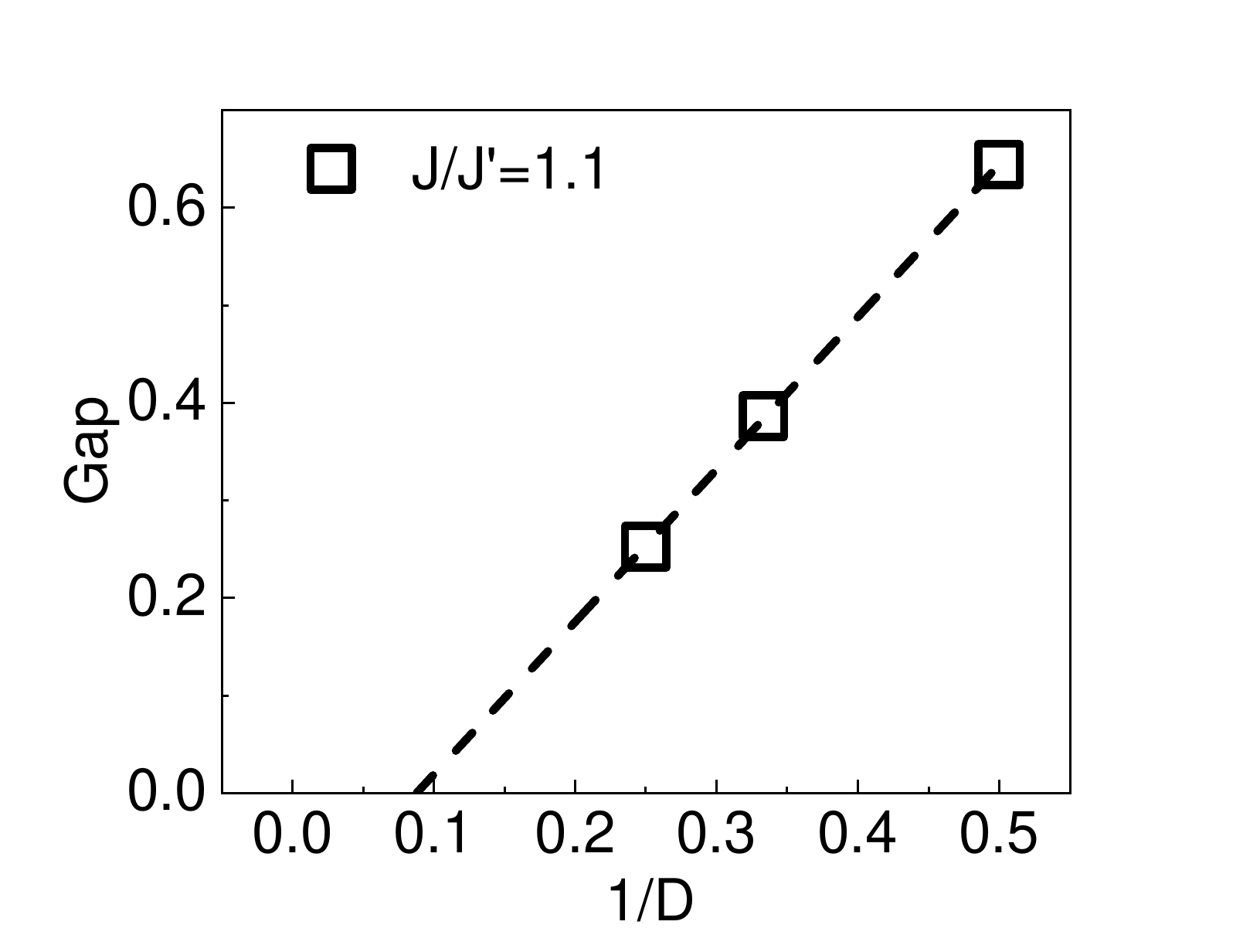}
		\caption{Finite $D$ scaling of the spin gap of the Goldstone modes (at M point) in the AFM phase of the SS model at $J/J^\prime=1.1$. The dashed line shows a best fit to the gap, which vanishes in the large $D$ limit.   }\label{fig:S3}
	\end{figure}
	
	Here we show numerical results on the spin excitation spectra in additional to those in the main text.
	The spectra illustrate the spin DSF
	\begin{equation}
	\mathcal{S}^{\alpha\beta}(\mathbf{q},\omega) = \frac{1}{N} \sum_{i,j}
	\int dt e^{i\mathbf{q}\cdot(\mathbf{r}_i-\mathbf{r}_j)} e^{i\omega t}
	\langle
	S_i^\alpha (t) S_j^\beta (0) \rangle
	\end{equation}
	with spin components $\alpha, \beta$.
	In Fig.~2(b) we have shown the spectra in the AFM phase at $J/J^\prime=1.1$ 
	calculated in the tensor network approach with bond dimension $D=3$. In 
	Fig.~\ref{fig:S2}, we show calculated spectra for several other $J/J^\prime$ 
	values and with different bond dimension $D$. One sees that the spectra show 
	similar features with two split magnon bands, which are the Goldstone modes, 
	at low energy and several longitudinal modes at high energy. Interestingly, 
	the spin gap of the Higgs mode
	decreases with decreasing $J/J^\prime$ toward the PVBS-to-AFM transition, 
	as discussed in the main text.
	
	As an effect of finite $D$, the Goldstone modes exhibit a finite gap. We have calculated the spin gaps with other $D$ values. As shown in Fig.~\ref{fig:S3}, the spin gap reduces with increasing $D$ and eventually vanishes in the large $D$ limit. This indicates that the gapless feature of Goldstone modes is captured by the tensor network calculation in this limit.
	
	\begin{figure}[!h]
		\centering
		\includegraphics[width=1\linewidth]{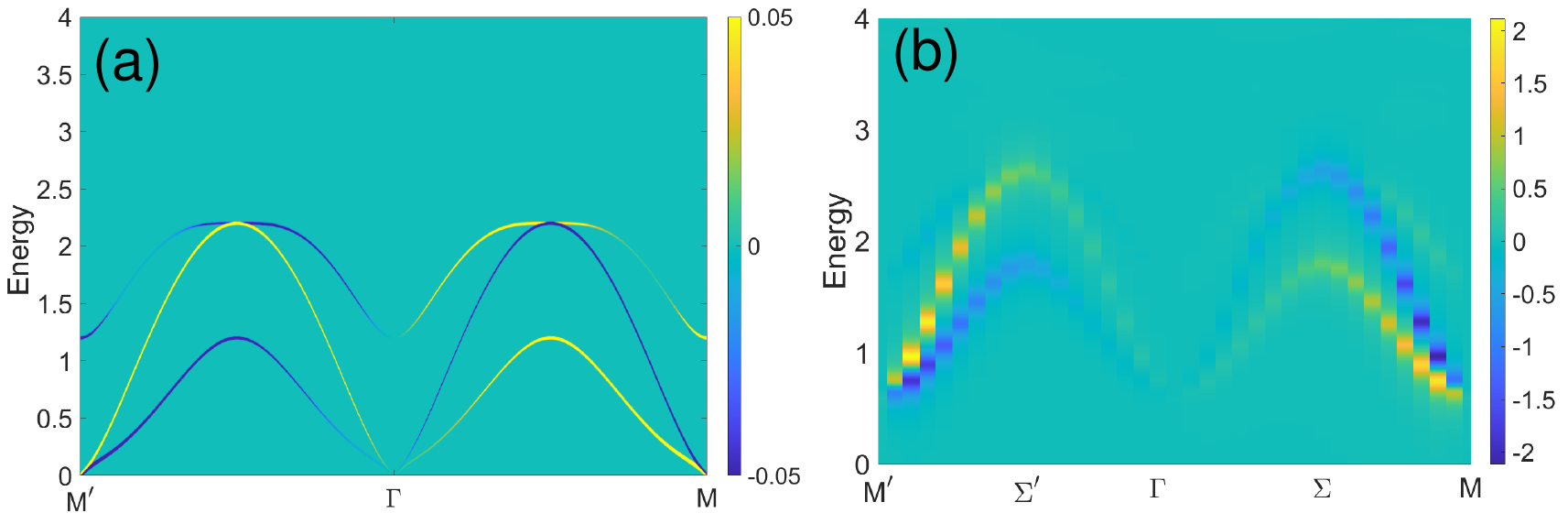}
		\caption{(a): Difference of the transverse DSF $\Delta \mathcal{S}= \mathcal{S}^{+-}(\mathbf{q},\omega) - \mathcal{S}^{-+}(\mathbf{q},\omega)$ in the AFM phase of the SS model at $J/J^\prime=1.1$ in the LSW theory. (b): Same as (a), but calculated by the tensor network method. In either case, the opposite sign of $\Delta\mathcal{S}$ reflects the opposite chirality of the magnon modes.}\label{fig:S4}
	\end{figure}
	
	To show that the chiral magnon excitations in the AFM phase of the SS model, we calculate the difference of the transverse component of the DSF $\Delta \mathcal{S}= \mathcal{S}^{+-}(\mathbf{q},\omega) - \mathcal{S}^{-+}(\mathbf{q},\omega)$. Note that $S_i^{+} S_j^{-} - S_i^{-} S_j^{+} \propto (\mathbf{S}_i\times \mathbf{S}_j)_z$, so that the sign of $\Delta \mathcal{S}$ is able to detect the chirality of the spin excitations. In Fig.~\ref{fig:S4} we compare results of $\Delta \mathcal{S}$ calculated by LSW theory and tensor network method. Both results show a finite splitting between the two Goldstone modes with opposite sign of $\Delta \mathcal{S}$. Also note that the magnon bands along $\Gamma$-M and $\Gamma$-M$^\prime$ also have opposite sign.
	Interestingly, we find that the optical magnon bands also exhibit opposite chirality.
	These results verify the existence of chiral magnons in the AFM state, which is caused by the symmetry of altermagnetism.
	
	\begin{figure}[!h]
		\centering
		\includegraphics[width=1\linewidth]{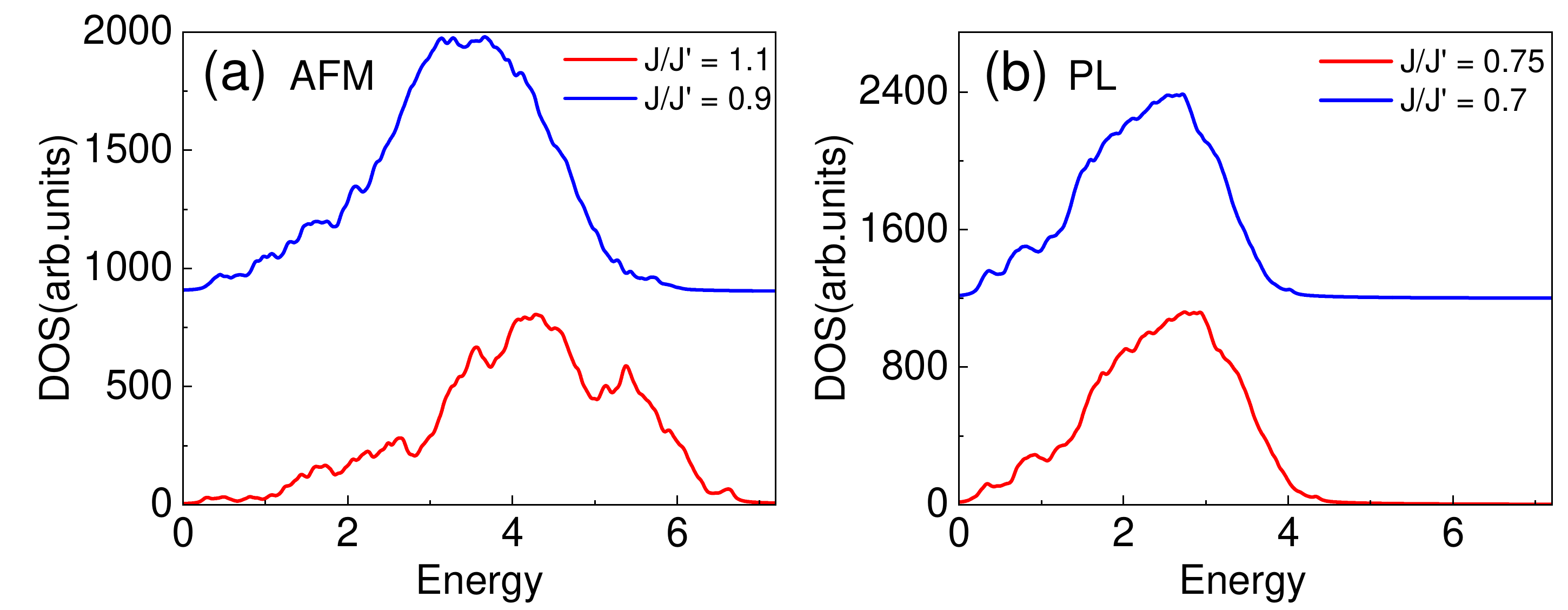}
		\caption{(a): Density of states (DoS) with excitation energy at several $J/J^\prime$ values in the AFM phase. (b): Same as (a) but in the plaquette phase.}\label{fig:S5}
	\end{figure}
	
	Besides the collective magnon excitations and the longitudinal modes, the 
	tensor calculation also resolves continuum excitations. The spectral 
	weights of these excitations are much smaller than the collective modes but 
	as shown in Fig.~\ref{fig:S5}, they cause a broad peak in the density of 
	states (DoS) at high energy. In the AFM phase, this broad peak appears in 
	energies above the Higgs gap (above $\sim 2J^\prime$ at $J/J^\prime=1.1$), 
	and the peak energy decreases with varying $J/J^\prime$ toward the 
	PVBS-to-AFM transition. This behavior is consistent with the emergence of 
	deconfined excitations although it is hard to tell whether the continuum 
	has a spinon or magnon origin. In the plaquette phase, we can also identify 
	continuous excitations at energies above the triplet gap (above $\sim 
	J^\prime$ at $J/J^\prime=0.7$) in the DoS plots (Fig.~\ref{fig:S5}(b)). The 
	behavior of energies of these excitations is similar to that in the AFM 
	phase.
	
	\begin{figure}[!h]
		\centering
		\includegraphics[width=1\linewidth]{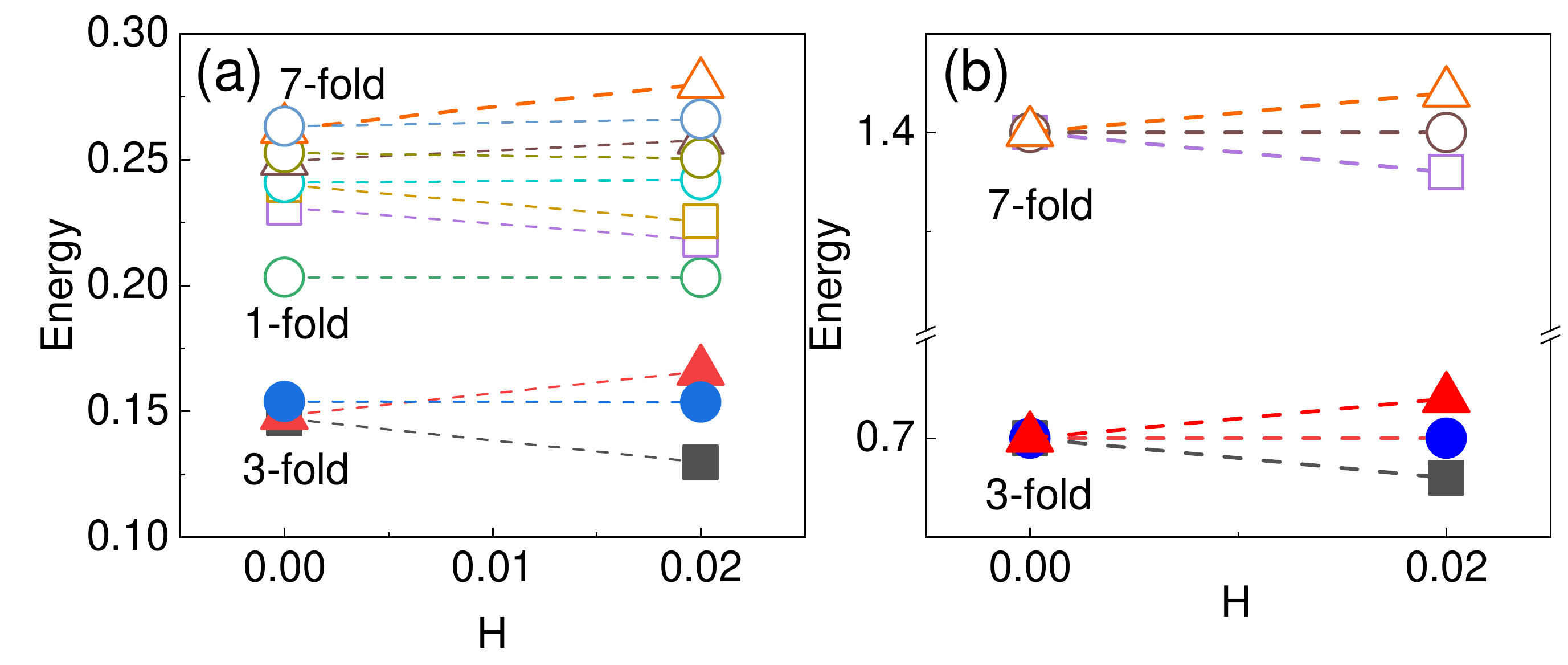}
		\caption{(a): Field dependence of gaps of several low-energy excitations 
		in the plaquette phase of the SS model, calculated by the tensor network 
		method. The multiplets split into $S^z=+1, -1, 0$ states. (b): Field 
		dependence of low-energy excitations of a 4-spin cluster with nearest 
		neighbor AFM interactions, calculated by the ED. The splitting indicates 
		that the 7-fold multiplet contains two degenerate triplets and one 
		singlet.
		}\label{fig:S6}
	\end{figure}
	
	To understand the singlet and the 7-fold multiplet modes above the lowest triplet mode in the plaquette phase resolved by the tensor network calculation in the main text, we apply a small magnetic field to split the multiplets. We also compare our tensor network results with those from exact diagonalization on a four-spin plaquette with nearest neighbor Heisenberg interaction. As shown in Fig.~\ref{fig:S6}(a), the 7-fold quasi-degenerate excitations splits under a small field. The gaps of three modes are unchanged under the field, while the gaps of two modes increase with the field, gaps of another two modes decrease with the field. Note that the slops of the gaps are the same for these modes, and they are also same to the slops of the split lowest triplet excitations. These indicate that the 7-fold multiplet actually consists of one singlet and two degenerate triplets. This conclusion is further confirmed by our exact diagonalization calculation shown in Fig.~\ref{fig:S6}(b). At zero field, there are indeed 7-fold degenerate multiplet excitations above the triplet ones. With applying a small magnetic field, The 7-fold multiplet splits into a pair of triplets and one singlet. Note that the singlet excitation in between the lowest triplet and the 7-fold multiplet seen in the tensor network result in Fig.~3 of the main text does not show up in our 4-spin cluster exact diagonalization calculation. It is then interpreted as an inter-plaquette bound state.

	\begin{figure}[!h]
		\centering
		\includegraphics[width=1\linewidth]{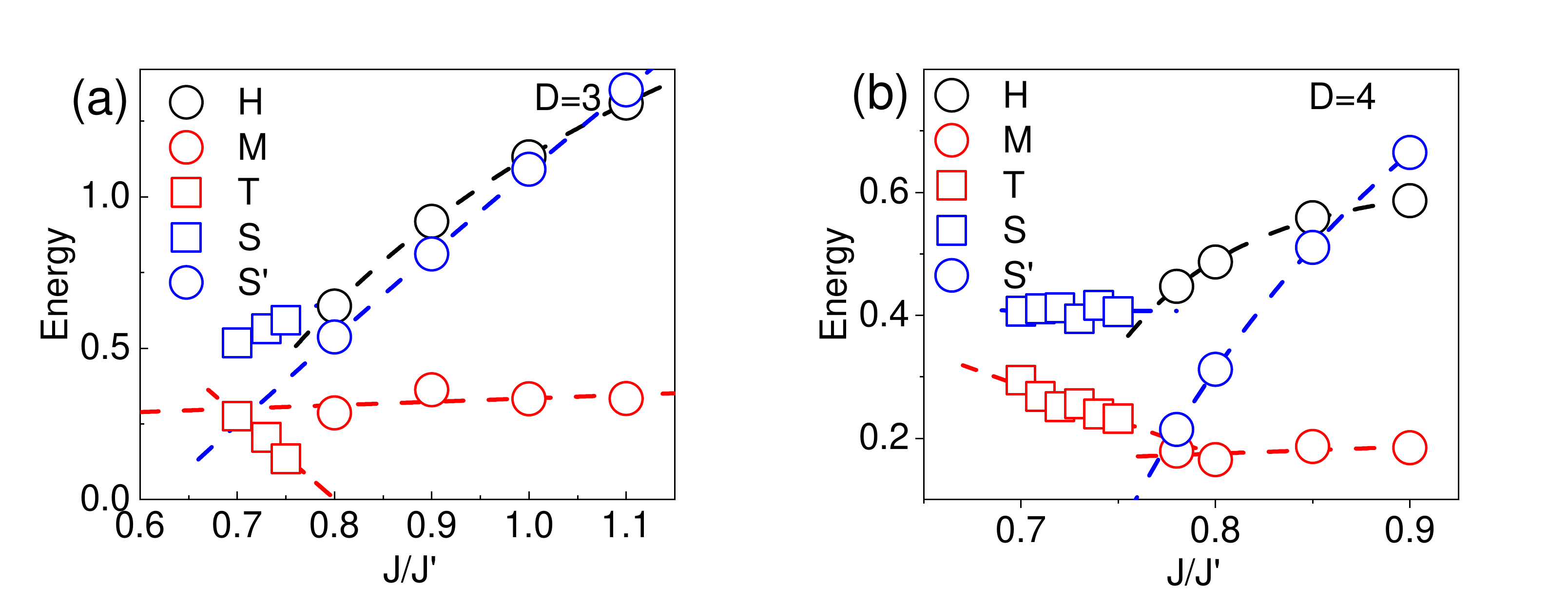}
		\caption{\ry{Evolution of gaps at the M point of the triplet (T) and 
		singlet (S) modes in the PVBS phase, as well as the Higgs (H), magnon 
		(M), and singlet (S$^\prime$) modes in the AFM phase, with
			$J/J^\prime$ near the PVBS-to-AFM transition in the iPEPS 
			calculation
			with $D=3$ (in (a)) and $D=4$ (in (b)). In either case we observe 
			softening of all collective modes when 
			approaching the transition point.}}\label{fig:S7}
	\end{figure}
	
	\ry{In Fig.~4(a) of the main text we discuss the evolution of various 
	collective excitations with varying $J/J^\prime$. These include triplet and 
	singlet modes in the PVBS phase, and  
	the Higgs, magnon, and singlet modes in the AFM phase. It is found 
	in Fig.~4(c) that 
	the gaps of triplet and singlet modes in the PVBS phase, together with 
	those of the Higgs and singlet modes in the AFM phase, drop rapidly to 
	small values close to zero, and become quasi-degenerate with the magnon 
	(Goldstone) modes when 
	approaching the PVBS-to-AFM transition. Here in Fig.~\ref{fig:S7} we 
	show results with smaller bond dimensions $D=3$ and $D=4$. We observe the 
	same behavior that all these collective modes are 
	softened near the transition. Note that the gap between the Higgs and 
	magnon modes is still sizable at the transition point for $D=3$ and $D=4$. 
	But as shown in Fig.~4(c) and (e), this gap drops rapidly with increasing 
	$D$ and eventually closes in the large $D$ limit. This behavior strongly 
	suggests the emergence of a DQCP at the PVBS-to-AFM transition.}

	 \bibliographystyle{unsrt}
	\bibliography{main}

\begin{thebibliography}{10}

\bibitem{TI_Review}
M.~Z. Hasan and C.~L. Kane.
\newblock Colloquium: topological insulators.
\newblock {\em Rev. Mod. Phys.}, 82(4):3045--3067, 2010.

\bibitem{Senthil_AnnRevCondMattPhys_2015}
T.~Senthil.
\newblock Symmetry-protected topological phases of quantum matter.
\newblock {\em Annu. Rev. Condens. Matter Phys.}, 6(1):299--324, 2015.

\bibitem{Witten_NP_2018}
E.~Witten.
\newblock Symmetry and emergence.
\newblock {\em Nat. Phys.}, 14(2):116--119, 2018.

\bibitem{Smejkal_PRX_2022A}
L.~{\v{S}}mejkal, J.~Sinova, and T.~Jungwirth.
\newblock Emerging research landscape of altermagnetism.
\newblock {\em Phys. Rev. X.}, 12(4):040501, 2022.

\bibitem{Smejkal_PRX_2022B}
L.~{\v{S}}mejkal, J.~Sinova, and T.~Jungwirth.
\newblock Beyond conventional ferromagnetism and antiferromagnetism: A phase
  with nonrelativistic spin and crystal rotation symmetry.
\newblock {\em Phys. Rev. X.}, 12(3):031042, 2022.

\bibitem{Yang_arXiv:2105.12738}
J.~Yang, Z.~Liu, and C.~Fang.
\newblock Symmetry invariants and classes of quasi-particles in magnetically
  ordered systems having weak spin-orbit coupling.
\newblock {\em arXiv:2105.12738}, 2021.

\bibitem{Smejkal_NatRevMater7_2022}
L.~{\v{S}}mejkal, A.~H. MacDonald, J.~Sinova, S.~Nakatsuji, and T.~Jungwirth.
\newblock Anomalous hall antiferromagnets.
\newblock {\em Nat. Rev. Mater.}, 7(6):482--496, 2022.

\bibitem{Gonzalez_Betancourt_PRL_2023}
R.~D. Gonzalez~Betancourt, J.~Zub{\'a}{\v{c}}, R.~Gonzalez-Hernandez,
  K.~Geishendorf, Z.~{\v{S}}ob{\'a}{\v{n}}, G.~Springholz, K.~Olejn{\'\i}k,
  L.~{\v{S}}mejkal, J.~Sinova, T.~Jungwirth, et~al.
\newblock Spontaneous anomalous hall effect arising from an unconventional
  compensated magnetic phase in a semiconductor.
\newblock {\em Phys. Rev. Lett.}, 130(3):036702, 2023.

\bibitem{Hariki_PRL_2024}
A.~Hariki, D.~A. Dal, O.~J. Amin, T.~Yamaguchi, A.~Badura, D.~Kriegner, K.~W.
  Edmonds, R.~P. Campion, P.~Wadley, D.~Backes, et~al.
\newblock X-ray magnetic circular dichroism in altermagnetic $\alpha$-{MnTe}.
\newblock {\em Phys. Rev. Lett.}, 132(17):176701, 2024.

\bibitem{Ferrari_arXiv:2408.00841}
F.~Ferrari and R.~Valenti.
\newblock Altermagnetism on the shastry-sutherland lattice.
\newblock {\em arXiv:2408.00841}, 2024.

\bibitem{Wu_PRB_2007}
Congjun Wu, Kai Sun, Eduardo Fradkin, and Shou-Cheng Zhang.
\newblock Fermi liquid instabilities in the spin channel.
\newblock {\em Phys. Rev. B}, 75:115103, Mar 2007.

\bibitem{Liu_NC_2021}
Hai-Yang Ma, Mengli Hu, Nana Li, Jianpeng Li, Wang Yao, Jin-Feng Jia, and
  Junwei Liu.
\newblock Multifunctional antiferromagnetic materials with giant piezomagnetism
  and noncollinear spin current.
\newblock {\em Nat. Commun.}, 12:2846, May 2021.

\bibitem{Satoru_JPSJ_2019}
Satoru Hayami, Yuki Yanagi, and Hiroaki Kusunose.
\newblock Momentum-dependent spin splitting by collinear antiferromagnetic
  ordering.
\newblock {\em Journal of the Physical Society of Japan}, 88(12):123702, 2019.

\bibitem{Smejkal_PRL131_256703_2023}
L.~{\v{S}}mejkal, A.~Marmodoro, K.~Ahn, R.~Gonz{\'a}lez-Hern{\'a}ndez,
  I.~Turek, S.~Mankovsky, H.~Ebert, S.~W. D’Souza, O.~{\v{S}}ipr, J.~Sinova,
  et~al.
\newblock Chiral magnons in altermagnetic ruo 2.
\newblock {\em Phys. Rev. Lett.}, 131(25):256703, 2023.

\bibitem{Ma_PRB110_064426_2024}
H.~Ma and J.~Jia.
\newblock Altermagnetic topological insulator and the selection rules.
\newblock {\em Phys. Rev. B.}, 110(6):064426, 2024.

\bibitem{Weissenhofer_PRB110_094427_2024}
M.~Wei{\ss}enhofer and A.~Marmodoro.
\newblock Atomistic spin dynamics simulations of magnonic spin seebeck and spin
  nernst effects in altermagnets.
\newblock {\em Phys. Rev. B.}, 110(9):094427, 2024.

\bibitem{Yao_arXiv_2024}
M.~Zhang, L.~Xiao, and D.~Yao.
\newblock Topological magnons in a collinear altermagnet.
\newblock {\em arXiv:2407.18379}, 2024.

\bibitem{Zhaojize_arxiv}
Y.~Liu, S.~Shao, S.~He, Z.~Y. Xie, J.~Mei, H.~Luo, and J.~Zhao.
\newblock Quantum dynamics in a spin-1/2 square lattice
  ${J_{1}}$-${J_{2}}$-$\delta$ altermagnet.
\newblock {\em arXiv:2410.06955}, 2024.

\bibitem{sachdev_2011}
S.~Sachdev and B.~Keimer.
\newblock Quantum criticality.
\newblock {\em Phys. Today}, 64(2):29--35, 2011.

\bibitem{Coldea_2010}
R.~Coldea, D.~A. Tennant, E.~M. Wheeler, E.~Wawrzynska, D.~Prabhakaran,
  M.~Telling, K.~Habicht, P.~Smeibidl, and K.~Kiefer.
\newblock Quantum criticality in an ising chain: experimental evidence for
  emergent ${E8}$ symmetry.
\newblock {\em Science}, 327(5962):177--180, 2010.

\bibitem{Senthil2004}
T.~Senthil, A.~Vishwanath, L.~Balents, S.~Sachdev, and M.~P.~A. Fisher.
\newblock Deconfined quantum critical points.
\newblock {\em Science}, 303(5663):1490--1494, 2004.

\bibitem{Cui2023DQCP}
Y.~Cui, L.~Liu, H.~Lin, K.~Wu, W.~Hong, X.~Liu, C.~Li, Z.~Hu, N.~Xi, S.~Li,
  et~al.
\newblock Proximate deconfined quantum critical point in
  {SrCu$_2$(BO$_3$)$_2$}.
\newblock {\em Science}, 380(6650):1179--1184, 2023.

\bibitem{Cui2019SCVO}
Y.~Cui, H.~Zou, N.~Xi, Z.~He, Y.X. Yang, L.~Shu, G.H. Zhang, Z.~Hu, T.~Chen,
  R.~Yu, et~al.
\newblock Quantum criticality of the ising-like screw chain antiferromagnet
  {SrCo$_{2}$V$_{2}$O$_{8}$} in a transverse magnetic field.
\newblock {\em Phys. Rev. Lett.}, 123(6):067203, 2019.

\bibitem{Lee2019}
J.~Y. Lee, Y.~Z. You, S.~Sachdev, and A.~Vishwanath.
\newblock {Signatures of a Deconfined Phase Transition on the
  Shastry-Sutherland Lattice: Applications to Quantum Critical
  SrCu$_2$(BO$_3$)$_2$}.
\newblock {\em Phys. Rev. X.}, 9, 2019.

\bibitem{E8_2021}
H.~Zou, Y.~Cui, X.~Wang, Z.~Zhang, J.~Yang, G.~Xu, A.~Okutani, M.~Hagiwara,
  M.~Matsuda, G.~Wang, et~al.
\newblock ${E8}$ spectra of quasi-one-dimensional antiferromagnet
  {BaCo$_2$V$_2$O$_8$} under transverse field.
\newblock {\em Phys. Rev. Lett.}, 127(7):077201, 2021.

\bibitem{Xu2022}
Y.~Xu, L.~S. Wang, Y.~Y. Huang, J.~M. Ni, C.~C. Zhao, Y.~F. Dai, B.~Y. Pan,
  X.~C. Hong, P.~Chauhan, S.~M. Koohpayeh, et~al.
\newblock Quantum critical magnetic excitations in spin-1/2 and spin-1 chain
  systems.
\newblock {\em Phys. Rev. X.}, 12(2):021020, 2022.

\bibitem{agrawal2020quantum}
U.~Agrawal, S.~Gopalakrishnan, and R.~Vasseur.
\newblock Quantum criticality in the 2d quasiperiodic potts model.
\newblock {\em Phys. Rev. Lett.}, 125(26):265702, 2020.

\bibitem{fey2019quantum}
S.~Fey, S.~C. Kapfer, and K.~P. Schmidt.
\newblock Quantum criticality of two-dimensional quantum magnets with
  long-range interactions.
\newblock {\em Phys. Rev. Lett.}, 122(1):017203, 2019.

\bibitem{zhang2011exploring}
X.~Zhang, C.~Hung, S.~Tung, N.~Gemelke, and C.~Chin.
\newblock Exploring quantum criticality based on ultracold atoms in optical
  lattices.
\newblock {\em New J. Phys.}, 13(4):045011, 2011.

\bibitem{Sandvik2007}
A.~W. Sandvik.
\newblock Evidence for deconfined quantum criticality in a two-dimensional
  heisenberg model with four-spin interactions.
\newblock {\em Phys. Rev. Lett.}, 98:227202, 2007.

\bibitem{Meng_EasyplaneDQCP}
N.~Ma, G.~Sun, Y.~You, C.~Xu, A.~Vishwanath, A.~W. Sandvik, and Z.~Meng.
\newblock Dynamical signature of fractionalization at a deconfined quantum
  critical point.
\newblock {\em Phys. Rev. B.}, 98(17):174421, 2018.

\bibitem{Shao_Science}
H.~Shao, W.~Guo, and A.~W. Sandvik.
\newblock Quantum criticality with two length scales.
\newblock {\em Science}, 352(6282):213--216, 2016.

\bibitem{Senthil_Vojta_Sachdev:2004}
T.~Senthil, M.~Vojta, and S.~Sachdev.
\newblock Weak magnetism and non-fermi liquids near heavy-fermion critical
  points.
\newblock {\em Phys. Rev. B.}, 69(3):035111, 2004.

\bibitem{Liu_Vojta:2022}
Z.~H. Liu, M.~Vojta, F.~F. Assaad, and L.~Janssen.
\newblock Metallic and deconfined quantum criticality in dirac systems.
\newblock {\em Phys. Rev. Lett.}, 128(8):087201, 2022.

\bibitem{Guo_PRL_2024}
Z.~Deng, L.~Liu, W.~Guo, and H.~Lin.
\newblock Diagnosing quantum phase transition order and deconfined criticality
  via entanglement entropy.
\newblock {\em Phys. Rev. Lett.}, 133(10):100402, 2024.

\bibitem{Shastry1981}
B.~{S. Shastry} and B.~Sutherland.
\newblock Exact ground state of a quantum mechanical antiferromagnet.
\newblock {\em Physica B+C}, 108(1):1069--1070, 1981.

\bibitem{Koga2000}
A.~Koga and N.~Kawakami.
\newblock {Quantum Phase Transitions in the Shastry-Sutherland Model for
  ${\mathrm{SrCu}}_{2}({\mathrm{BO}}_{3}{)}_{2}$}.
\newblock {\em Phys. Rev. Lett.}, 84:4461--4464, 2000.

\bibitem{Chung2001}
C.~H. Chung, J.~B. Marston, and S.~Sachdev.
\newblock Quantum phases of the shastry-sutherland antiferromagnet: Application
  to (formula presented).
\newblock {\em Phys. Rev. B.}, 64, 2001.

\bibitem{Pixley2014}
J.~H. Pixley, R.~Yu, and Q.~Si.
\newblock Quantum phases of the shastry-sutherland kondo lattice: Implications
  for the global phase diagram of heavy-fermion metals.
\newblock {\em Phys. Rev. Lett.}, 113:176402, 2014.

\bibitem{Corboz2013}
P.~Corboz and F.~Mila.
\newblock Tensor network study of the shastry-sutherland model in zero magnetic
  field.
\newblock {\em Phys. Rev. B.}, 87:115144, 2013.

\bibitem{Boos2019}
C.~Boos, S.~P.~G. Crone, I.~A. Niesen, P.~Corboz, K.~P. Schmidt, and F.~Mila.
\newblock {Competition between intermediate plaquette phases in
  ${\mathrm{SrCu}}_{2}({\mathrm{BO}}_{3}){}_{2}$ under pressure}.
\newblock {\em Phys. Rev. B.}, 100, 2019.

\bibitem{yang2022quantum}
J.~Yang, A.~W. Sandvik, and L.~Wang.
\newblock Quantum criticality and spin liquid phase in the shastry-sutherland
  model.
\newblock {\em Phys. Rev. B.}, 105(6):L060409, 2022.

\bibitem{Sengupta_PRL:2013}
K.~Wierschem and P.~Sengupta.
\newblock Columnar antiferromagnetic order and spin supersolid phase on the
  extended shastry-sutherland lattice.
\newblock {\em Phys. Rev. Lett.}, 110(20):207207, 2013.

\bibitem{Wang_PRL_2023}
J.~Wang, H.~Li, N.~Xi, Y.~Gao, Q.~Yan, W.~Li, and G.~Su.
\newblock Plaquette singlet transition, magnetic barocaloric effect, and spin
  supersolidity in the shastry-sutherland model.
\newblock {\em Phys. Rev. Lett.}, 131(11):116702, 2023.

\bibitem{Xi2023}
N.~Xi, H.~Chen, Z.~Y. Xie, and R.~Yu.
\newblock Plaquette valence bond solid to antiferromagnet transition and
  deconfined quantum critical point of the shastry-sutherland model.
\newblock {\em Phys. Rev. B.}, 107(22):L220408, 2023.

\bibitem{Corboz_arXiv_2025}
Philippe Corboz, Yining Zhang, Boris Ponsioen, and Fr\'{e}d\'{e}ric Mila.
\newblock Quantum spin liquid phase in the shastry-sutherland model revealed by
  high-precision infinite projected entangled-pair states.
\newblock {\em arXiv:2502.14091}, 2025.

\bibitem{WenyuanLiu_PRL_2024}
W.~Liu, X.~Zhang, Z.~Wang, S.~Gong, W.~Chen, and Z.~Gu.
\newblock Quantum criticality with emergent symmetry in the extended
  shastry-sutherland model.
\newblock {\em Phys. Rev. Lett.}, 133(2):026502, 2024.

\bibitem{Mila_BOT}
M.~Moliner, I.~Rousochatzakis, and F.~Mila.
\newblock Emergence of one-dimensional physics from the distorted
  shastry-sutherland lattice.
\newblock {\em Phys. Rev. B.}, 83(14):140414, 2011.

\bibitem{Pinaki_PRB92_094440_2015}
Z.~Zhang and P.~Sengupta.
\newblock Generalized plaquette state in the anisotropic shastry-sutherland
  model.
\newblock {\em Phys. Rev. B.}, 92(9):094440, 2015.

\bibitem{WangZ2018}
Z.~Wang and C.~D. Batista.
\newblock Dynamics and instabilities of the shastry-sutherland model.
\newblock {\em Phys. Rev. Lett.}, 120(24):247201, 2018.

\bibitem{Wang2023}
K.~Liu and F.~Wang.
\newblock Schwinger boson symmetric spin liquids of shastry-sutherland model.
\newblock {\em Phys. Rev. B.}, 109(13):134409, 2024.

\bibitem{Kageyama1999}
H.~Kageyama, K.~Yoshimura, R.~Stern, N.~V. Mushnikov, K.~Onizuka, M.~Kato,
  K.~Kosuge, C.~P. Slichter, T.~Goto, and Y.~Ueda.
\newblock Exact dimer ground state and quantized magnetization plateaus in the
  two-dimensional spin system ${\mathrm{srcu}}_{2}({\mathrm{bo}}_{3}){}_{2}$.
\newblock {\em Phys. Rev. Lett.}, 82:3168--3171, 1999.

\bibitem{PhysRevLett.82.3701}
S.~Miyahara and K.~Ueda.
\newblock {Exact Dimer Ground State of the Two Dimensional Heisenberg Spin
  System ${\mathrm{SrCu}}_{2}({\mathrm{BO}}_{3}){}_{2}$}.
\newblock {\em Phys. Rev. Lett.}, 82:3701--3704, 1999.

\bibitem{Waki2007}
T.~Waki, K.~Arai, M.~Takigawa, Y.~Saiga, Y.~Uwatoko, H.~Kageyama, and Y.~Ueda.
\newblock {A novel ordered phase in SrCu$_2$ (BO$_3$)$_2$ under high pressure}.
\newblock {\em J. Phys. Soc. Jpn.}, 76, 2007.

\bibitem{haravifard2016crystallization}
S.~Haravifard, D.~Graf, A.~E. Feiguin, C.~D. Batista, J.~C. Lang, D.~M.
  Silevitch, G.~Srajer, B.~D. Gaulin, H.~A. Dabkowska, and T.~F. Rosenbaum.
\newblock {Crystallization of spin superlattices with pressure and field in the
  layered magnet ${\mathrm{SrCu}}_{2}({\mathrm{BO}}_{3}{)}_{2}$}.
\newblock {\em Nat Commun}, 7(1):1--6, 2016.

\bibitem{Zayed2017}
M.~E. Zayed, C.~R\"uegg, J.~Larrea, A.~M. L\"auchli, C.~Panagopoulos, S.~S.
  Saxena, M.~Ellerby, D.~F. Mcmorrow, Th~Str\"assle, S.~Klotz, et~al.
\newblock {4-spin plaquette singlet state in the Shastry-Sutherland compound
  SrCu$_2$ (BO$_3$)$_2$}.
\newblock {\em Nat. Phys}, 13:962--966, 2017.

\bibitem{Bettler2020}
S.~Bettler, L.~Stoppel, Z.~Yan, S.~Gvasaliya, and A.~Zheludev.
\newblock Sign switching of dimer correlations in
  ${\mathrm{srcu}}_{2}{({\mathrm{BO}}_{3})}_{2}$ under hydrostatic pressure.
\newblock {\em Phys. Rev. Res.}, 2:012010, 2020.

\bibitem{Guo2020}
J.~Guo, G.~Sun, B.~Zhao, L.~Wang, W.~Hong, V.A. Sidorov, N.~Ma, Q.~Wu, S.~Li,
  Z.~Meng, A.W. Sandvik, and L.~Sun.
\newblock {Quantum Phases of ${\mathrm{SrCu}}_{2}({\mathrm{BO}}_{3}{)}_{2}$
  from High-Pressure Thermodynamics}.
\newblock {\em Phys. Rev. Lett.}, 124:206602, 2020.

\bibitem{Jimenez2021}
J.~L. Jim\'enez, S.~P.~G. Crone, E.~Fogh, M.~E. Zayed, R.~Lortz,
  E.~Pomjakushina, K.~Conder, A.~M. L\"euchli, L.~Weber, S.~Wessel, and et~al.
\newblock A quantum magnetic analogue to the critical point of water.
\newblock {\em Nature}, 592(7854):370--375, 2021.

\bibitem{verstraete2004renormalization}
F.~Verstraete and J.~I. Cirac.
\newblock Renormalization algorithms for quantum-many body systems in two and
  higher dimensions.
\newblock {\em arXiv:cond-mat/0407066}, 2004.

\bibitem{orus2014practical}
R.~Or{\'u}s.
\newblock A practical introduction to tensor networks: Matrix product states
  and projected entangled pair states.
\newblock {\em Ann. Phys.}, 349:117--158, 2014.

\bibitem{corboz2014competing}
P.~Corboz, T.~M. Rice, and M.~Troyer.
\newblock Competing states in the t-j model: Uniform d-wave state versus stripe
  state.
\newblock {\em Phys. Rev. Lett.}, 113(4):046402, 2014.

\bibitem{orus2009simulation}
R.~Or{\'u}s and G.~Vidal.
\newblock Simulation of two-dimensional quantum systems on an infinite lattice
  revisited: Corner transfer matrix for tensor contraction.
\newblock {\em Phys. Rev. B.}, 80(9):094403, 2009.

\bibitem{PhysRevB.84.041108}
P.~Corboz, S.~R. White, G.~Vidal, and M.~Troyer.
\newblock Stripes in the two-dimensional $t$-$j$ model with infinite projected
  entangled-pair states.
\newblock {\em Phys. Rev. B.}, 84:041108, 2011.

\bibitem{SRGAD}
B.~Chen, Y.~Gao, Y.~Guo, Y.~Liu, H.~Zhao, H.~Liao, L.~Wang, T.~Xiang, W.~Li,
  and Z.~Y. Xie.
\newblock Automatic differentiation for second renormalization of tensor
  networks.
\newblock {\em Phys. Rev. B.}, 101:220409, 2020.

\bibitem{liao2019differentiable}
H.~Liao, J.~Liu, L.~Wang, and T.~Xiang.
\newblock Differentiable programming tensor networks.
\newblock {\em Phys. Rev. X}, 9(3):031041, 2019.

\bibitem{ponsioen2020excitations}
B.~Ponsioen and P.~Corboz.
\newblock Excitations with projected entangled pair states using the corner
  transfer matrix method.
\newblock {\em Phys. Rev. B.}, 101(19):195109, 2020.

\bibitem{ponsioen2022automatic}
B.~Ponsioen, F.~F. Assaad, and P.~Corboz.
\newblock Automatic differentiation applied to excitations with projected
  entangled pair states.
\newblock {\em SciPost Phys.}, 12(1):006, 2022.

\bibitem{chi2022spin}
R.~Chi, Y.~Liu, Y.~Wan, H.~Liao, and T.~Xiang.
\newblock Spin excitation spectra of anisotropic spin-1/2 triangular lattice
  heisenberg antiferromagnets.
\newblock {\em Phys. Rev. Lett.}, 129(22):227201, 2022.

\bibitem{Poldolsky_PRB84_174522_2011}
D.~Podolsky, A.~Auerbach, and D.~P. Arovas.
\newblock Visibility of the amplitude (higgs) mode in condensed matter.
\newblock {\em Phys. Rev. B.}, 84(17):174522, 2011.

\bibitem{SM}
See supplemental material \url{http://link...} for details about the tensor
  network method, the linear spin wave theory, the bond-operator theory, and
  additional results on the spin excitation spectra, which also includes
  refs.\cite{verstraete2004renormalization,orus2014practical,
  corboz2014competing,orus2009simulation,PhysRevB.84.041108,SRGAD,
  liao2019differentiable,ponsioen2020excitations,ponsioen2022automatic,chi2022spin,VCTMRG2022,LBFGS,zhitomirsky1996valence,sachdev1990bond}.

\bibitem{Song_NC_2019}
Xue-Yang Song, Chong Wang, Ashvin Wishwanath, and Yin-Chen He.
\newblock Unifying description of competing orders in two-dimensional quantum
  magnets.
\newblock {\em Nat. Commun.}, 10:4254, Sep 2019.

\bibitem{Maity_arXiv_2025}
Atanu Maity, Francesco Ferrari, Jong~Yeon Lee, Janik Potten, Tobias M\"{u}ller,
  Ronny Thomale, Rhine Samajdar, and Yasir Iqbal.
\newblock Evidence for a $\textrm{Z}_2$ dirac spin liquid in the generalized
  shastry-sutherland model.
\newblock {\em arXiv:2501.00096}, 2025.

\bibitem{Zhao_NP_2018}
B.~Zhao, P.~Weinberg, and A.~W. Sandvik.
\newblock Symmetry-enhanced discontinuous phase transition in a two-dimensional
  quantum magnet.
\newblock {\em Nat. Phys.}, 15(7):678--682, 2019.

\bibitem{ChengchenLi_JPCM_2024}
C.~Li, H.~Lin, and R.~Yu.
\newblock Quantum scaling of the spin lattice relaxation rate in the
  checkerboard jq model.
\newblock {\em J. Phys.: Condens. Matter}, 36(35):355805, 2024.

\bibitem{Zhao_PRL_2020}
Bowen Zhao, Jun Takahashi, and Anders~W. Sandvik.
\newblock Multicritical deconfined quantum criticality and lifshitz point of a
  helical valence-bond phase.
\newblock {\em Phys. Rev. Lett.}, 125:257204, Dec 2020.

\bibitem{Lu_PRB_2021}
Da-Chuan Lu, Cenke Xu, and Yi-Zhuang You.
\newblock Self-duality protected multicriticality in deconfined quantum phase
  transitions.
\newblock {\em Phys. Rev. B}, 104:205142, Nov 2021.

\bibitem{ma2020theory}
R.~Ma and C.~Wang.
\newblock Theory of deconfined pseudocriticality.
\newblock {\em Phys. Rev. B.}, 102(2):020407, 2020.

\bibitem{nahum2020note}
A.~Nahum.
\newblock Note on wess-zumino-witten models and quasiuniversality in 2+ 1
  dimensions.
\newblock {\em Phys. Rev. B.}, 102(20):201116, 2020.

\bibitem{zhou2024so}
Z.~Zhou, L.~Hu, W.~Zhu, and Y.~He.
\newblock {SO}(5) deconfined phase transition under the fuzzy-sphere
  microscope: Approximate conformal symmetry, pseudo-criticality, and operator
  spectrum.
\newblock {\em Phys. Rev. X.}, 14(2):021044, 2024.

\bibitem{LilingSun_arXiv_2024}
J.~Guo, P.~Wang, C.~Huang, B.~Chen, W.~Hong, S.~Cai, J.~Zhao, J.~Han, X.~Chen,
  Y.~Zhou, et~al.
\newblock Deconfined quantum critical point lost in pressurized srcu2 (bo3) 2.
\newblock {\em arXiv:2310.20128}, 2023.

\bibitem{Cui_PRB_2023}
Q.~Cui, B.~Zeng, P.~Cui, T.~Yu, and H.~Yang.
\newblock Efficient spin seebeck and spin nernst effects of magnons in
  altermagnets.
\newblock {\em Phys. Rev. B.}, 108(18):L180401, 2023.

\bibitem{VCTMRG2022}
X.~F. Liu, Y.~F. Fu, W.~Q. Yu, J.~F. Yu, and Z.~Y. Xie.
\newblock Variational corner transfer matrix renormalization group method for
  classical statistical models.
\newblock {\em Chin. Phys. Lett}, 39(6):067502, 2022.

\bibitem{LBFGS}
J.~Nocedal and S.~J. Wright.
\newblock {\em Numerical Optimization}.
\newblock Springer New York, NY, 3 edition, 2006.

\bibitem{sachdev1990bond}
S.~Sachdev and R.~N. Bhatt.
\newblock Bond-operator representation of quantum spins: Mean-field theory of
  frustrated quantum heisenberg antiferromagnets.
\newblock {\em Phys. Rev. B.}, 41(13):9323, 1990.

\bibitem{zhitomirsky1996valence}
M.~E. Zhitomirsky and K.~Ueda.
\newblock Valence-bond crystal phase of a frustrated spin-1/2 square-lattice
  antiferromagnet.
\newblock {\em Phys. Rev. B.}, 54(13):9007, 1996.

\end{thebibliography}
\end{document}